\documentclass[journal]{IEEEtran}
\usepackage{algorithm}
\usepackage[justification=centering]{caption}
\usepackage{algpseudocode}
\usepackage{fancyhdr}
\usepackage{textcomp}
\usepackage{amsmath}
\usepackage{subfigure}
\usepackage{amssymb,amsfonts}
\usepackage{graphicx}
\usepackage{textcomp}
\usepackage{xcolor}
\usepackage{multicol}
\usepackage{multirow}
\usepackage{blindtext}
\usepackage{adjustbox}

\algnewcommand{\Initialize}[1]{%
  \State \textbf{Initialize:}
\parbox[t]{.8\linewidth}{\raggedright #1}
}
\algnewcommand{\Goto}{\textbf{go to}}%

\def\BibTeX{{\rm B\kern-.05em{\sc i\kern-.025em b}\kern-.08em
    T\kern-.1667em\lower.7ex\hbox{E}\kern-.125emX}}

\usepackage{graphicx}
\usepackage{ulem}

\fancypagestyle{firstpage}{
    \fancyhf{} 
    \lhead{This paper has been accepted by IEEE Transactions on Communications}
}

\begin{document}

\title{Mobile Traffic Prediction using LLMs with Efficient In-context Demonstration Selection}

\author{Han~Zhang,
        Akram Bin Sediq,
        Ali Afana,
        and~Melike~Erol-Kantarci,~\IEEEmembership{Fellow, IEEE}
\thanks{Han Zhang and Melike Erol-Kantarci are with the School of Electrical Engineering and Computer Science, University of Ottawa, Ottawa, ON K1N 6N5, Canada (e-mail: hzhan363@uottawa.ca; melike.erolkantarci@uottawa.ca).}
\thanks{Akram Bin Sediq and Ali Afana are with Ericsson, Ottawa, K2K 2V6, Canada(e-mail:
akram.bin.sediq@ericsson.com; ali.afana@ericsson.com)
}}

\markboth{ }%
{Shell \MakeLowercase{\textit{et al.}}: Bare Demo of IEEEtran.cls for IEEE Journals}

\maketitle
\thispagestyle{firstpage}

\begin{abstract}
Mobile traffic prediction is an important enabler for optimizing resource allocation and improving energy efficiency in mobile wireless networks. Building on the advanced contextual understanding and generative capabilities of large language models (LLMs), this work introduces a context-aware wireless traffic prediction framework powered by LLMs. To further enhance prediction accuracy, we leverage in-context learning (ICL) and develop a novel two-step demonstration selection strategy, optimizing the performance of LLM-based predictions.
The initial step involves selecting ICL demonstrations using the effectiveness rule, followed by a second step that determines whether the chosen demonstrations should be utilized, based on the informativeness rule. We also provide an analytical framework
for both informativeness and effectiveness rules. The effectiveness of the proposed framework is demonstrated with a real-world fifth-generation (5G) dataset with different application scenarios. According to the numerical results, the proposed framework shows lower mean squared error and higher $R^2$-Scores compared to the zero-shot prediction method and other demonstration selection methods, such as constant ICL demonstration selection and distance-only-based ICL demonstration selection.

\end{abstract}

%
\IEEEpeerreviewmaketitle

\section{Introduction}
In the past few years, the wireless mobile traffic data have continued to grow due to breakthroughs in wireless access technologies such as network densification, large reflecting surfaces, and terahertz spectrum \cite{zhou2024heuristic}\cite{chen2021intelligent}. The explosive growth in demand for data transmission is placing severe requirements on the upcoming sixth-generation (6G) wireless networks \cite{jiang2021road}.

Mobile traffic prediction is the process of predicting the future downlink traffic bitrate
of every single mobile device using current and historical traffic data along with contextual information. It is a significant capability to enhance wireless network management, enabling optimized resource allocation strategies and improved energy efficiency \cite{ArafatTraffic}. 
This also helps to mitigate network traffic congestion and ensure enhanced quality of service and quality of experience \cite{10625144}. Many research efforts have already been dedicated to making accurate wireless traffic predictions. Some popular approaches include using moving average models \cite{kapgate2014weighted} and deep learning (DL) models for prediction \cite{zhang2021dual}. However, real-world wireless network traffic data are highly nonlinear and complex, especially when the data types are constantly diversifying \cite{iturria2023rl}. Therefore, accurate traffic prediction remains a challenging problem.

Recently, large language models (LLMs) have attracted much attention in various areas given their impressive capacity for information comprehension and strategic planning \cite{habib2024llm}. This capability is largely attributed to the attention mechanism, which allows LLMs to capture long-range dependencies and contextual relationships more effectively \cite{woo2018cbam,ouyang2023efficient,yu2024phase}. Increasing numbers of papers have used LLMs for time series forecasting and it has been demonstrated that pre-trained LLMs can lead to good performance in some time series analysis tasks \cite{zhou2023one}. However, currently, there are only a limited number of works that use LLMs for mobile traffic prediction. 

LLMs offer key advantages for mobile traffic prediction over traditional machine learning models \cite{zhang2024large}. Their generative capabilities, enhanced by in-context learning (ICL), enable a one-size-fits-all approach for traffic prediction \cite{gruver2024large}, eliminating the need for retraining or adaptation to different network conditions. In contrast, traditional ML models usually require task-specific training and fine-tuning. LLMs also exhibit greater robustness to dynamic network conditions since they can rapidly adapt to evolving traffic patterns using in-context examples without parameter modification. In contrast, traditional ML models often require frequent retraining when the network conditions change. This adaptability makes LLMs suitable for scenarios with fluctuations due to user mobility, network congestion, and environmental changes. Moreover, mobile traffic prediction is related to contextual factors such as device velocity, position, signal-to-interference-and-noise ratio, and radio access technologies. Traditional ML models usually require explicit data pre-processing and feature engineering to convert input into numerical formats. In contrast, LLMs can directly comprehend and integrate raw, heterogeneous contextual information. This eliminates the need for manual data transformation and preserves the original meaning of these factors.

Despite the benefits of LLM-based mobile traffic prediction, it still suffers from limitations such as instability and hallucinations. In-context learning (ICL) is an effective method of guiding LLM responses by including known examples or demonstrations in the prompts. Compared with fine-tuning methods, ICL only relies on the prompt design, and the parameters of LLM remain unchanged during ICL \cite{zhang2024large}. In this sense, ICL provides more flexibility and efficiency, since it can be performed across various tasks without re-training the pre-trained LLMs, and less data and computational resources are required. Therefore, the ICL method is applied in our proposed LLM-based mobile traffic prediction framework to improve the quality of LLM outputs.

Yet, the performance of ICL suffers from instability. It is highly related to the selection and permutation of the demonstrations, and the optimal selection method usually varies between tasks. Moreover, according to existing research, ICL exhibits different behaviors between larger models and smaller models. Specifically, larger models are more sensitive to the unrelated information in the ICL demonstrations and can be easily distracted \cite{shi2023large}\cite{shi2024largerlanguagemodelsincontext}. This indicates that in some cases, the misuse of ICL can even negatively impact the LLM performance. Therefore, with the advent of the era of large models, it is crucial to perform ICL properly and to select good demonstrations for ICL.

Building upon these insights, we propose an LLM-enabled mobile traffic prediction framework with a two-step ICL demonstration selection scheme. This work differs from existing works in the following aspects. First, some previous LLM-based time series prediction works usually used smooth and periodic time series data, which simplifies the prediction task \cite{gruver2024large}. In contrast, our work utilizes real-world fifth-generation (5G) traffic data \cite{raca2020beyond} with complex data pattern as test input. This ensures the effectiveness and applicability of the proposed algorithm in practical wireless communication scenarios. Second, existing LLM-based traffic prediction frameworks have rarely considered the integration of ICL \cite{zhou2023one}. The few studies using ICL have not discussed how to perform ICL demonstration selection, which limits the effectiveness of ICL. Moreover, most of the existing ICL demonstration selection methods are intuition-based or heuristic, lacking a robust theoretical foundation \cite{qin2023context}. Unlike previous studies, our approach provides a more systematic and effective method for selecting ICL demonstrations.

The main contributions of this work are listed as follows:

1) We propose an LLM-enabled context-aware wireless traffic prediction framework. In addition, we perform the ICL algorithm and integrate selected historical demonstrations in the prompt as the source of long-term memory to enhance prediction accuracy and reduce hallucinations. 

2) We design a two-step efficient ICL demonstration selection algorithm to decide when to use ICL and how to find good ICL demonstrations in the given task. Specifically, we propose two rules to evaluate the ICL demonstrations, i.e., effectiveness and informativeness, and develop methods to quantify these two metrics. Our proposed ICL demonstration selection algorithm can bring three key advantages. First, it is developed on the basis of a detailed theoretical analysis of ICL, ensuring that the selection process follows well-defined principles. Second, unlike some ICL approaches that require a large number of ICL demonstrations to achieve meaningful improvements, our method demonstrates that even one or two carefully chosen demonstrations can significantly enhance model performance. By selecting only the most impactful demonstrations, our method improves the prediction accuracy while minimizing token usage. Furthermore, by leveraging the most effective demonstrations, it enables smaller-scale LLMs with limited reasoning ability to benefit from ICL.

3) The superiority of the proposed framework is demonstrated on a real-world 5G dataset. We also demonstrate that ICL-based LLMs offer greater generalization compared to custom-built machine learning models. These models can effectively leverage demonstrations from diverse scenarios to improve prediction accuracy, even when the test data originates from a different context. This adaptability minimizes the reliance on scenario-specific data and retraining, enhancing the resource efficiency and flexibility of LLMs. Our selection strategy is performed on a per-sample basis and it is tested under a wide range of conditions, including different types of network activities and both driving and static settings. The same LLM can be used to make accurate predictions under diverse and evolving traffic patterns without any updates.

In the context of mobile communication, the implications of this work are threefold: First, traffic prediction serves as a foundation for proactive network control. As a result, our work has direct implications for real-time network optimization and predictive analysis. Our ICL demonstration selection method can be extended beyond mobile traffic prediction and is applicable to various LLM-driven tasks in wireless networks. Finally, our method eliminates the need for fine-tuning or training the model, making it highly suitable for resource-constrained mobile communication systems with dynamic network conditions.

The rest of the paper is organized as follows. Section II introduces related works. Section III formulates the traffic prediction problem and defines the LLM-empowered mobile traffic prediction framework. Section IV describes the implementation of the two-step efficient demonstration selection method. Section V explains experimental settings and experimental results, and Section VI concludes the paper.

\section{Related Works}
There have been many studies that discuss how to accurately predict mobile traffic. Several works have used DL to learn traffic patterns from historical data. For instance, \cite{mei2022realtime} developed a DL model for real-time downlink bandwidth prediction and handover prediction in a 5G communication network. In \cite{he2020meta}, a meta-learning scheme was proposed to perform short-term predictions of user traffic intensity. \cite{sun2021mobile} proposed a graph-based temporal convolutional network to predict the future traffic of each network unit in a wireless network. In \cite{trinh2018mobile}, the long short-term memory model (LSTM) was applied to the one-step wireless traffic prediction. With the advent of transformer-based architectures, several existing works have explored their use for traffic prediction. For instance, \cite{ArafatTraffic} used a transformer architecture to predict wireless network traffic in concise temporal intervals for open radio access networks. \cite{liu2021st} designed transformer blocks to model the temporal and spatial features of traffic flows in communication networks. Some other works performed traffic prediction based on moving average methods. For instance, \cite{dalmazo2017performance} proposed a weighted moving average (WMA) model that can estimate the traffic of the next interval. In \cite{tian2021network}, an auto-regressive integrated moving average (ARIMA) model was used to predict the network traffic time series. Different from these works, our paper proposes a pre-trained LLM-based mobile traffic prediction framework. Unlike DL-based and transformer-based methods, our proposed framework eliminates the need for scenario-specific model training, addressing challenges including the data scarcity and the computational costs of training. Pre-trained LLMs can be directly applied in a one-size-fits-all manner. Compared with moving average methods, our proposed framework has a higher accuracy, especially for nonlinear time series.

Some other existing works leveraged LLMs for time series prediction. For example, \cite{zhou2023one} performed thorough experiments and has shown that pre-trained language models can achieve comparable performance in major types of time series analysis tasks. However, these studies did not focus on mobile traffic prediction problems. Mobile traffic prediction is inherently different from other time series tasks due to its complex data patterns and the need to incorporate contextual network conditions. These factors make some previously proposed methods inapplicable to our task.

Some other existing studies fine-tuned LLMs or trained LLMs from scratch for time series prediction. For instance, \cite{cao2023tempo} leveraged an interpretable prompt-tuning-based generative transformer to learn time series representations. \cite{chang2023llm4ts} proposed two fine-tuning strategies to align LLMs with the nuances of time-series data. It has been shown that fine-tuning can outperform ICL-based method and other prompt engineering techniques in certain settings. However, it usually requires a large amount of task-specific labeled data, whereas ICL can operate effectively with limited data. Moreover, traffic patterns often evolve over time, which means frequent model updates are required by fine-tuning. This leads to additional computational costs, energy costs and maintenance costs. Another limitation is that fine-tuning may cause the model to overfit to specific traffic scenarios, resulting in degraded performance when the scenario changes. Ultimately, the choice between fine-tuning and ICL should be guided by the requirements and constraints of the target application, including data availability, resource limits, and deployment flexibility.

\cite{hu2024self} is the only existing work that targets predicting wireless network traffic with LLMs. In this work, a self-refined LLM was designed to refine incorrect predictions through a three-step process. The key difference lies in the ICL demonstration selection method. \cite{hu2024self} focuses on predicting hourly traffic within a single day, where the data follows simpler patterns and available ICL demonstrations are relatively similar. As a result, a random selection of ICL examples does not significantly impact performance. In contrast, our work deals with more complex traffic patterns and emphasizes improving prediction accuracy by selecting high-quality ICL demonstrations rather than relying on random selection.

Some other works discussed what makes good ICL demonstrations. In \cite{xie2021explanation}, the ICL process was explained as implicit Bayesian inference by extracting a latent concept and modeling a hidden Markov model. \cite{han2023context} demonstrated that the ICL process has some similarities with the kernel regression. \cite{dai2023can} explained LLMs as meta-optimizers and understands ICL as implicit fine-tuning. These works serve as theoretical bases for us to analyze how to select ICL demonstrations. 

\begin{figure*}[t]
\centering
\includegraphics[width=6.5in]{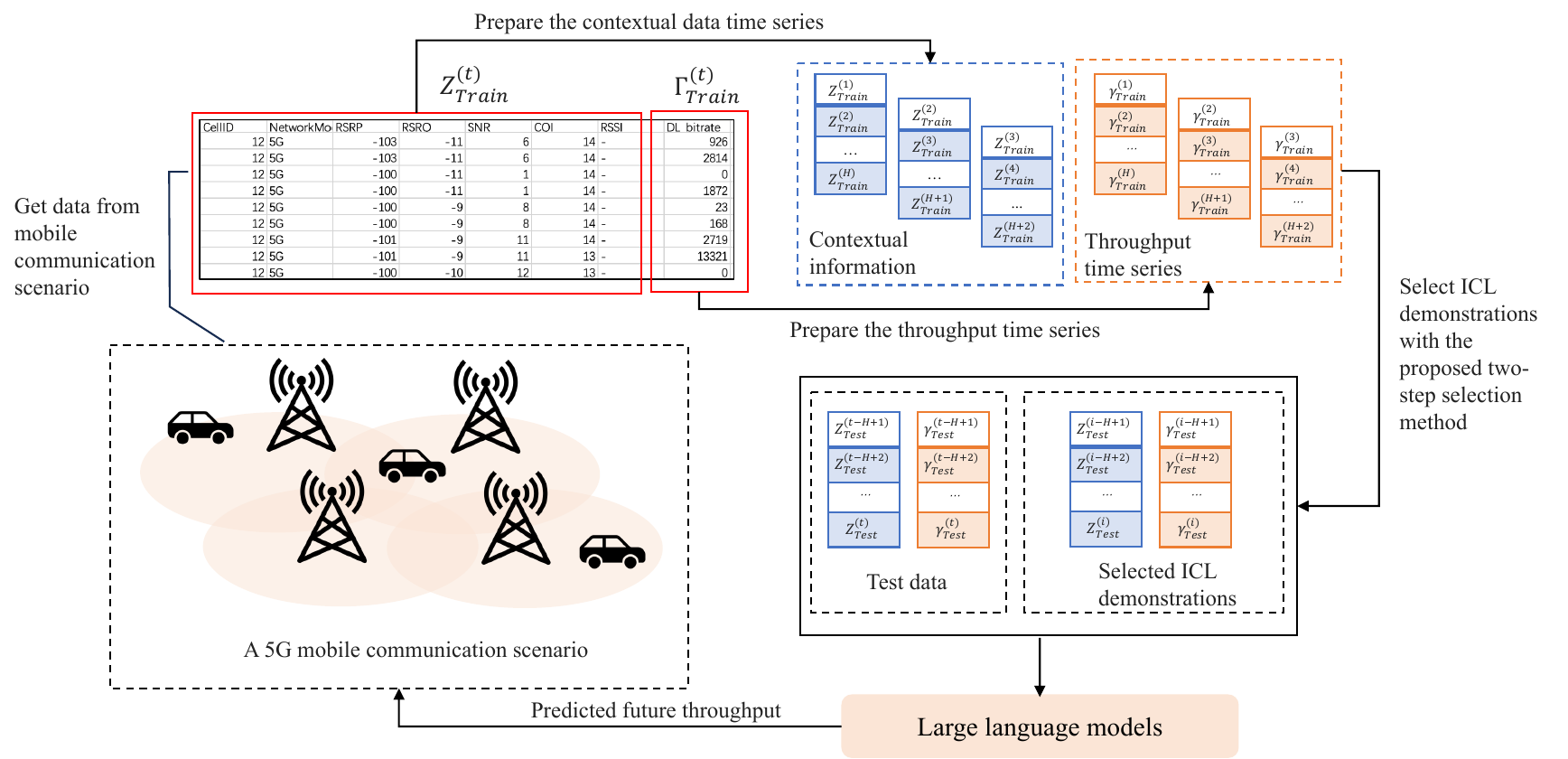}
\caption{System design of LLM-based mobile traffic prediction.}
\label{fig1}
\end{figure*}

Moreover, \cite{rubin2021learning} proposed to train an efficient dense retriever to retrieve training examples as prompts. \cite{zhang2023makes} proposed both supervised and unsupervised prompt retrieval methods to choose examples for visual ICL. \cite{qin2023context} proposed an iterative demonstration selection method and iteratively selected examples that are strongly correlated with the test sample. Most of these methods were tested with classification tasks and may not be suitable for time-series forecasting tasks. Moreover, they are mostly intuition-based or heuristic methods and lack theoretical support. Unlike these approaches, our proposed framework is derived from a fine-grained theoretical analysis. Through the analysis, we also offer a clear interpretation of the intrinsic relationships between the LLM output and selected ICL demonstrations.

\section{System Design}

Fig. \ref{fig1} shows the system design of our proposed LLM-based mobile traffic prediction framework. In this framework, data are first collected from a 5G mobile communication scenario and screened with a two-step ICL demonstration selection scheme. The selected ICL examples are then combined with the test data as input to the LLM for future traffic prediction. In this section, we first define the system model and formulate the traffic prediction problem. Next, we explain how to pre-process data for the ICL demonstration setup and how to generate prompts. 

\subsection{Problem Formulation}

This study aims to predict the future downlink traffic bitrate of every single mobile device using current and historical traffic data along with contextual information. The base station (BS) can periodically measure the bitrate over the communication channel and create discrete time series representing the device's traffic patterns. Additionally, the BS also has access to some contextual data, including the channel metrics and the neighboring cell metrics, to enhance the accuracy of the prediction \cite{yue2017linkforecast}.

Due to the limited size of the prompts, only part of the historical information is used as the input for mobile traffic prediction. The historical time window $H$ is defined to control the input length. In this case, at time $t$, the input downlink throughput time series $\Gamma^{(t)}$ can be formulated as:
\begin{align}
    \Gamma^{(t)} = \{\gamma^{(t)}, \gamma^{(t-1)}, ..., \gamma^{(t-H+1)}\}.
\end{align}
The input contextual information $Z^{(t)}$ can be formulated as:
\begin{align}
    Z^{(t)} = 
\left(
\begin{matrix}
z_{1}^{(t)} & z_{2}^{(t)} & \cdots & z_{N}^{(t)} \\
z_{1}^{(t-1)} & z_{2}^{(t-1)} & \cdots & z_{N}^{(t-1)} \\
\vdots & \vdots & \vdots & \vdots \\
z_{1}^{(t-H+1)} & z_{2}^{(t-H+1)} & \cdots & z_{N}^{(t-H+1)} \\
\end{matrix}
\right),
\end{align}
where $\{z_1, z_2, ..., z_{N}\}$ denotes the $N$ input features selected from all the available contextual information. In this scenario, we focus only on the LLM-enabled one-step ahead traffic prediction problem and the prediction process can be formulated as:
\begin{align}
    \hat{y}^{(t)} = f(\Gamma^{(t)},Z^{(t)}|\theta^{pre}),\label{eq-3}
\end{align}
where $\hat{y}^{(t)}$ denotes the predicted future downlink throughput. $f(.)$ denotes the prediction model and $\theta^{pre}$ denotes the model parameters of pre-trained LLMs. 

On this basis, the ICL method is applied to improve the accuracy of prediction. ICL is a prompt engineering method for LLMs to learn new tasks by including known examples in the prompt. It presents the advantage of using off-the-shelf LLMs to solve new tasks without explicit re-training and fine-tuning. However, the performance of ICL can be unstable and highly affected by the demonstration selection \cite{zhang2022active}. As a result, the selection method of ICL demonstrations is important to improve the accuracy of traffic prediction. To distinguish between the data used to develop the ICL demonstrations and the data used for testing, the ICL demonstrations are assumed to be selected from a training dataset where the true future throughput values are known. The test data are from another testing dataset with no overlap between the two datasets. To perform a $M$-shot prediction, the index set of selected ICL demonstrations $I^{(t)}$ is defined as:
\begin{align}
    I^{(t)} = \{i_1,i_2, ... i_M\},
\end{align}
where $M$ denotes the total number of selected ICL demonstrations. The selected ICL demonstrations set are then defined as:
\begin{align}
    S^{(t)} = \{[(\Gamma_{tr}^{(i)}, Z_{tr}^{(i)}),y_{tr}^{(i)}]|\forall i \in I^{(t)}\},
\end{align}
where $(\Gamma_{tr}^{(i)}, Z_{tr}^{(i)})$ denotes the input downlink throughput time series and the input contextual information of the $i^{th}$ training data. $y_{tr}^{(i)}$ denotes the corresponding true future throughput value. Eq. (\ref{eq-3}) can be seen as the prediction process without ICL. By integrating the selected demonstrations into Eq. (\ref{eq-3}), the ICL-based traffic prediction can be then rewritten as:
\begin{align}
     \hat{y}^{(t)}_{te} = f([\Gamma_{te}^{(t)},Z_{te}^{(t)}],S^{(t)}|\theta^{pre}).
\end{align}
The optimization goal is to minimize the gap between the predicted downlink throughput and the true downlink throughput, which is given as:
\begin{align}
    \underset{I^{(t)}}{\text{argmin}}\ \frac{1}{T}\sum_{t}L(\hat{y}^{(t)}_{te},y^{(t)}_{te}),
\end{align}
where $y^{(t)}_{te}$ denotes the true future throughput value of the given test data. $T$ denotes the length of the test set. $L(\hat{y}^{(t)}_{te},y^{(t)}_{te})$ denotes the loss function and it can be defined as the squared Euclidean distance, which is given as:
\begin{align}
    L(\hat{y}_{te}^{(t)},y_{te}^{(t)}) = (\hat{y}_{te}^{(t)}-y^{(t)}_{te})^2
\end{align}

\subsection{Data Preprocessing and Prompt design}
To create ICL demonstrations and set up tests, the whole dataset is first divided into two parts. The first part is considered the training dataset, while the second part serves as the testing dataset. Next, both datasets are partitioned into multiple partially overlapping short sequences. The length of these sequences is equal to the historical time window $H$. 

During the prediction, the ICL sequences are first selected from the training dataset as context series. Then they are combined with the testing time series for prompt generation. In this way, the prediction is made by LLMs based on two parts of memories. 
The first part of memory is the long-term memory provided by the context series. It shows the historical traffic patterns of the mobile device and indicates some consistent information such as the maximum capacity of the communication channel and the impact of the contextual information on the traffic.
The second part of the memory is the short-term memory provided by the testing time series from the most recent previous period. It indicates the current trend of traffic flow.

Based on this design, the prompt of the mobile traffic prediction task includes four parts, task description, demonstrations, input information, and output formatting. The task description part focuses on describing the known information of the scenario and the role LLM plays in the task. For instance, the task description describes the traffic prediction task and indicates the coexistence of 5G and long-term evolution (LTE) technologies setup. 

\begin{figure*}[h]
\centering
\includegraphics[width=6.5in]{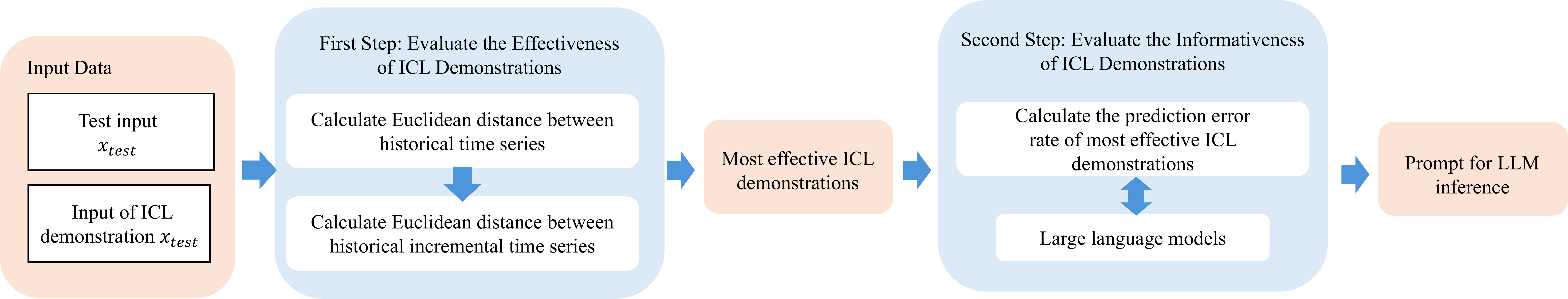}
\caption{The workflow of the two-step ICL demonstration selection method.}\label{fig2step}
\vspace{-15pt}
\end{figure*}
\vspace{20pt}

In the next part, the selected ICL examples are converted into human-like language descriptions by substituting numeric values into a natural language template. In case of missing data, the corresponding contextual descriptions are removed to ensure the coherence of the prompts. After conversion, the ICL examples are aligned and integrated into the prompt. Similarly, in the input information part, the test data are also converted into a natural language format and integrated into the prompt. Specifically, to limit the output length and enhance prediction stability, a mechanism is designed for the automatic conversion of the measurement unit of the downlink traffic. 

In our algorithm design, it is observed that the model becomes less accurate when predicting traffic with larger magnitudes. One possible reason is that LLMs generate predictions digit by digit. As the number of digits increases, the correlation between different digits may weaken, leading to reduced prediction accuracy and stability. To mitigate such concerns, when the throughput values in the historical traffic time series are greater than $10^5$, the throughput units are converted from kilobits per second to megabits per second, and the values are scaled down accordingly. With this mechanism, the precision of the predicted values can still reach 1\%, and the hallucinations and instability due to long digits are reduced.

Output formatting is also included to define how LLMs should output the results. With clear output formatting, the predicted values can be easily extracted from the natural language output.

\subsection{Deployment Considerations for LLM-based Predictions}

In this subsection, the deployment considerations for LLM-based predictions in real-world 5G and 6G networks are discussed. The LLM inference is intended to run on edge servers located at BSs or access points, or cloud servers. These servers are usually equipped with graphics processing units (GPUs) and AI accelerator hardware, which can support the large computation demand of LLM efficiently. Moreover, they also have powerful and fast-access storage resources to meet the storage requirements and low-latency inference requirements. On these infrastructure, a LLM with less than 15 billion parameters is expected to achieve an inference time of around 100 ms \cite{chitty2024llm}, with a network communication latency of around 50 ms \cite{fan2025latency}, depending on hardware settings and network conditions.

These delays remain well within acceptable limits for per-second traffic prediction. For real-world deployment in 5G or 6G networks, hybrid strategies can also be adopted to further ensure the feasibility of the proposed method. Lightweight LLMs can be deployed at the edge to handle predictions involving simpler traffic patterns, while larger LLMs in the cloud can be selectively invoked for more complex scenarios.

Another concern of large-scale deployment in 5G or 6G networks is that LLM inference usually consumes a lot of energy. Deploying LLM to edge or cloud servers helps minimize the energy impact on end-user devices and enables a better resource utilization. Furthermore, in real-world deployment, optimization techniques can be used to further reduce energy consumption. For example, model quantization strategies can compresses model weights to lower bit precision to the memory usage and energy usage of LLMs \cite{xiao2023smoothquant}. Similarly, KV caching reduces redundant computation by storing key and value tensors from previous tokens and this also reduce the energy consumption of LLM inference \cite{li2024does}.

In summary, the proposed framework is feasible for real-world deployment in 5G and 6G networks, with acceptable latency and manageable hardware requirements. Moreover, optimization techniques can be used to reduce the energy consumption of LLM in large-scale deployments.

\section{Two-step efficient demonstration selection}
In this section, we explain how we perform efficient demonstration selection for ICL-based traffic prediction. Specifically, we first discuss what makes a suitable demonstration for ICL that can improve the prediction accuracy as much as possible. Then, we introduce our proposed two-step selection method.

\subsection{Finding Suitable Demonstrations for ICL}
Although the effect of ICL has been validated in previous studies, the performance of ICL still suffers from instability. It is highly task-dependent and easily influenced by ICL demonstrations' selection.
To optimize the accuracy gains from integrating ICL while minimizing the computational costs, we first need to decide whether to apply ICL for each test case and then we select the ICL demonstrations that are most likely to improve the prediction accuracy.

An intuitive way to select demonstrations is to include all the available demonstrations in the prompt. Each new demonstration addition is expected to bring more usable information, thus improving the LLM performance.
However, in real-world tasks, the length of the prompt is usually restricted and only a minimal number of ICL demonstrations can be integrated into the prompt. In addition, an increased number of ICL examples also results in longer inference time and requires more memory to run. This will constrain the practical use of LLMs in many scenarios. Moreover, undesired ICL examples are of limited help in improving the LLM performance. In some complicated tasks, they may even provide redundant information and add extra difficulties for LLMs to understand the task \cite{guo2024makes}. Therefore, it is critical to decide 
whether to apply ICL for each test case and which ICL demonstrations to use.

In general, LLMs expect ICL demonstrations to provide four types of information, namely, the output format, the distribution of the input, the range of the output, and the correlation of inputs and outputs \cite{min2022rethinking}. In our system design, the output format has been indicated by the output formatting part in the prompt. The distribution of the input and the range of the output can also be inferred from the task description and the pre-trained knowledge base of LLMs. Therefore, in the given traffic prediction task, the ICL demonstration selection largely depends on whether the demonstration can provide valid information regarding the correlation of inputs and outputs.

Based on the above thoughts, we propose two rules that should be considered during the ICL demonstration selection: \textit{\textbf{effectiveness}} and \textit{\textbf{informativeness}}. Effectiveness means that the selected ICL demonstrations can provide valid information and are helpful for the test case. Informativeness means that the selected ICL examples can provide information that the original pre-trained LLM does not have. According to these two rules, in this work, we design a two-step ICL demonstration selection method, which is shown in Fig. \ref{fig2step}. In the first step, the similarity between ICL demonstrations and the test data are calculated by measuring the distance between historical time series and incremental time series. This step is performed to ensure the effectiveness of selected ICL examples. In the second step, the prediction error rate of the selected ICL demonstration is measured to evaluate the informativeness of the example and decide if the ICL should be used for the given test case. 

In the next subsection, a theoretical analysis of ICL is provided and detailed explanations about how to develop the two-step ICL demonstration selection method are given.

\subsection{First step: Evaluate the effectiveness of ICL demonstrations}

To understand the theory of ICL, the output generation of a pre-trained LLM is first modeled as a hidden Markov model \cite{xie2021explanation}. The general form can be given as:
\begin{equation}
o_j \sim P(o_j = o| s_j = s),
\end{equation}
where $o_j$ denotes the $j^{th}$ output token, and $P(.)$ denotes the probability. $s_j$ denotes the state before generating the $j^{th}$ output token, which is related to the input prompt and the previous state. This formulation means the next output character is sampled from a probability distribution of all the possible tokens according to the current state. The state transition can be formulated as:
\begin{equation}
s_j \sim  P(s_{j+1} = s'| s_j = s).  
\end{equation}
With this setting, the parameter $\theta$ can be used to illustrate the transition probabilities matrix of the hidden Markov process and it is developed from the prompts.

To evaluate the effectiveness of ICL demonstrations, following \cite{han2023context}, we first explain the similarity between the ICL and the kernel regression process. We define the prediction task as $\theta^{*}$, and the task $\theta^{*}$ describes the relationship between the input and the output of the prediction. During the prediction, LLMs first calculate the probabilities of all the output tokens. A kernel function can be constructed as:
\begin{equation}
    K(x,x') = <\hat{P}(o|x,\theta^{pre})^{T},\Sigma_{pre}^{-1}\hat{P}(o|x',\theta^{pre})>,
\end{equation}
where $x = [\Gamma^{(t)}, Z^{(t)}]$ denotes the input of LLMs in the case of zero-shot prediction. $<.,.>$ denotes the inner product. $\hat{P}(o|x,\theta^{pre})$ is a vector that denotes the probabilities of all the output tokens given the input $x$ under pre-trained LLM parameters $\theta^{pre}$. In the following of this subsection, we denote $\hat{P}(o|x,\theta^{pre})$ as $\hat{P}_x$ for brevity. $\Sigma_{pre}$ is the covariance of the $\hat{P}(o|x,\theta^{pre})$.

By leveraging the property of the kernel function, a series of equations can be constructed, which are then used to establish the relationship between ICL and kernel regression. According to the definition of covariance, the kernel function has the following properties:
\begin{equation}
    \begin{split}
    &\mathbb{E}_{x_{D}}[K(x_{te},x_{D})\hat{P}_{x_{D}}^T]
    \\= &\hat{P}_{x_{te}}^{T}\Sigma_{pre}^{-1}\mathbb{E}[\hat{P}_{x_{D}}\hat{P}_{x_{D}}^T]
    \\= &\hat{P}_{x_{te}}^T,        
    \end{split}
\end{equation}

where $x_{D} = [\Gamma_{tr}^{(i)}, Z_{tr}^{(i)}]$ denotes the input of selected ICL demonstrations. $x_{te} = [\Gamma_{te}^{(i)}, Z_{te}^{(i)}]$ denotes the input of test data. $\mathbb{E}_{x_{D}}[K(x_{te},x_{D})\hat{P}_{x_{D}}^T]$ denotes the expectations of $K(x_{te},x_{D})\hat{P}_{x_{D}}^T$ given $x_{D}$. Considering the linear correspondence between the two ends of the equation, the equation can be transformed as:
\begin{equation}
\begin{split}
    &\mathbb{E}_{x_{D}}[K(x_{te},x_{D})P(o = y_{te}|x_{D},\theta^{pre})] \\= &P(o = y_{te}|x_{te},\theta^{pre}).
\end{split}
\end{equation}

In kernel regression, if the selected ICL demonstrations are used as observed data points, the predicted values of test data $x_{te}$ can be estimated as:
\begin{equation}
\begin{split}
    \hat{y}_{k} =&  \frac{\sum^{M}_{m=1}e(y_{tr}^{(i_m)})K(x_{te},x_{D})}{\sum^{M}_{m=1}K(x_{te},x_{D})}
    \\=&\frac{1}{M}\sum^{M}_{m=1}\hat{P}(y|x^{(i_m)}_{te},\theta^{pre})^{T}\Sigma_{pre}^{-1}\hat{P}(y|x^{(i_m)}_{D},\theta^{pre})e(y_{D}^{(i_m)}),
\end{split}
\end{equation}
where $\hat{y}_{k}$ denotes the predicted values of kernel regression. $e(y_{D}^{(i_m)})$ denotes the one-hot vector for the output $y_{D}^{(i_m)}$. 

Since the task is described as $\theta^*$, the data pairs $(x^{(i_m)}_{D}, y_{D}^{(i_m)})$ can be seen as sampled from the probability distribution $P(y = Y|x^{(i_m)}_{D},\theta^*)$. According to Hoeffding’s inequality, the upper bound on the probability that the sum of random variables deviates from its expected value can be qualified. If the probability is set as $1-2\delta$, the inequality can be converted into:
\begin{equation}
\begin{split}
    &||\hat{y}_{k}-\mathbb{E}_{x_{D}}[K(x_{te},x_{D})P(o = y_{te}|x_{D},\theta^{*})]||_{\infty}
    \\=&||\frac{1}{M}\sum^{M}_{m=1}\hat{P}(y|x^{(i_m)}_{te},\theta^{*})^{T}\Sigma_{pre}^{-1}\hat{P}(y|x^{(i_m)}_{D},\theta^{*})e(y_{D}^{(i_m)})\\&-\mathbb{E}_{x_{D}}[K(x_{te},x_{D})P(o = y_{te}|x_{D},\theta^{*})]||_{\infty}
    \\< &\sqrt{\frac{1}{2M}\ln{\frac{4D}{\delta}}},
\end{split}
\end{equation}
where $||.||_{\infty}$ denotes the L-infinity norm. $D$ denotes the dimension of $\hat{P}(y|x^{(i_m)}_{te},\theta^{*})$. We denote the upper bound of Frobenius-norm of $\hat{P}(o|x,\theta)$ as $\eta$. Since the relationship between inputs and outputs of the traffic prediction task is partially consistent with the knowledge gained by the LLMs through pre-training, the difference between the output distribution under $\theta^{pre}$ and $\theta^{*}$ can be upper-bounded as:
\begin{equation}
\begin{split}
    &|<\hat{P}(y|x_{te},\theta^{*})^{T},\Sigma_{*}^{-1}\hat{P}(y|x_{D},\theta^{*})> - \\&<\hat{P}(y|x_{te},\theta^{pre})^{T},\Sigma_{pre}^{-1}\hat{P}(y|x_{D},\theta^{pre})>|\leq \eta^2\epsilon_\theta,
\end{split}
\end{equation}
where $\epsilon_\theta$ is defined to quantize the upper limit. Therefore, the relationship between $\hat{y}$ and $P(o = y_{te}|x_{te},\theta^{*})$ can be derived:
\begin{equation}
||\hat{y}-P(o = y_{te}|x_{te},\theta^{*})||_{\infty}  \leq \sqrt{\frac{1}{2m}\ln{\frac{4D}{\delta}}} + \eta^2\epsilon_\theta.
\end{equation}

The above inference process illustrates the high similarity between the estimates of $P(o = y_{te}|x_{te},\theta^{*})$ and kernel regression. According to \cite{dai2023can}, the behavior of ICL is similar to explicit fine-tuning with ICL demonstrations, which means $P(o = y_{te}|x_{te},\theta^{*})$ also has a high similarity with $P(o = y_{te}|x_{te},x_{D},\theta^{pre})$. Therefore, it can be concluded that the ICL process is comparable to kernel regression based on selected demonstrations. 

According to the property of kernel regression, demonstrations that are similar to the test data are supposed to lead to better estimation results. Therefore, the demonstrations can be selected by minimizing the kernel similarity between the test data and the selected ICL demonstration, which is given as:
\begin{equation}
\underset{i}{\min}\ [\hat{P}_{x_{te}}^{T}\Sigma_{pre}^{-1}\hat{P}_{x_{D}}].
\end{equation}
For the traffic prediction problem, it is difficult to compute the above values directly. However, the kernel similarity can be understood as the semantic similarity between the test input $x_{te}$ and input of ICL demonstration $x_{D}$. 
In this work, we measure the semantic similarity from two aspects. For the first aspect, the Euclidean distance between the historical downlink throughput time series of test data and the ICL demonstrations is calculated as:
\begin{equation}
    e_1 = ||\Gamma_{te}-\Gamma_{D}||_2.
\end{equation}
For the second aspect, the incremental time series of downlink throughput is first calculated:
\begin{equation}
    \Delta\Gamma^{(t)} = \{\gamma^{(t)} - \gamma^{(t-1)}, ..., \gamma^{(t-H+2)}-\gamma^{(t-H+1)}\}.
\end{equation}
Next, the Euclidean distance between the incremental time series of test data and the ICL demonstrations is calculated as:
\begin{equation}
    e_2 = ||\Delta\Gamma_{te}-\Delta\Gamma_{D}||_2.
\end{equation}
The sum of $e_1$ and $e_2$ is used to estimate the effectiveness of ICL demonstrations. A smaller sum indicates a higher similarity between the ICL demonstration and the test sample, suggesting greater effectiveness of the ICL demonstration. Therefore, we can select the most effective samples by minimizing the sum, which can be written as $\min\ (e_1+e_2)$.

In the following, a theoretical analysis is provided of how the first-step ICL demonstration selection impacts the LLM performance. Without any demonstrations, the LLM relies solely on its pre-trained knowledge to make predictions. The prediction error can be formulated as:
\begin{equation}
    Er(\hat{y_{test}}, y_{test}) = \epsilon_{noise} + \epsilon_{LLM}
    \label{eq22}
\end{equation}
where $Er(\hat{y_{test}}, y_{test})$ denotes the prediction error, $\epsilon_{noise}$ denotes the intrinsic noise error, and $\epsilon_{LLM}$ denotes the error due to LLM's lack of task-specific knowledge. Since ICL is theoretically related to kernel regression, after applying ICL, the error can be reformulated as:

\begin{equation}
    Er(\hat{y_{test}}, y_{test}) = \epsilon_{noise} + O(1-K(x_{test},x_{ICL}))
    \label{eq23}
\end{equation}

where $O(.)$ is the asymptotic upper bound of the error term. As shown in the formulation, $\epsilon_{LLM}$ was replaced by $O(1-K(x_{test},x_{ICL}))$ through ICL. That means ICL reduces prediction error when $K(x_{test},x_{ICL})$ is maximized.

In the first step of demonstration selection, we maximize $K(x_{test},x_{ICL})$ by selecting the demonstration most similar to $x_{test}$. This ensures that through ICL, the selected demonstration minimizes the prediction error introduced by the LLM's lack of task-specific knowledge. Moreover, compared with random or constant ICL demonstration selection methods, our proposed method ensures larger $K(x_{test},x_{ICL})$, thereby leading to greater stability in LLM-based predictions.

\subsection{Second step: Evaluate the informativeness of ICL demonstrations}
In this subsection, we explain the second step of ICL demonstration selection and illustrate how the informativeness of ICL demonstrations is evaluated. In this step, ICL will be performed only if the selected ICL demonstrations are informative, i.e., the selected ICL examples can provide information the original pre-trained LLM does not have. In this way, the computational and negative impact of unnecessary ICL can be reduced.

Following the hidden Markov model defined in the last subsection, the loss function of ICL under the set of selected demonstrations $S$ can first be defined as:
\begin{equation}
L(\theta;S,t) = \log P(y_{te}^{(t)}|\theta,S^{(t)},x_{te}^{(t)}),
\end{equation}
where $P(y_{te}|\theta, S,x_{te})$ denotes the probability of obtaining the desired output of the given test case under the selected demonstrations $S$. 

With this loss function, the optimization goal of ICL can be formulated as:
\begin{equation}
\begin{split}
\theta^{opt} &= \underset{\theta}{\text{argmin}}L(\theta;S,t) \\&= \underset{\theta}{\text{argmin}}(-\log P(y^{(t)}_{te}|\theta,S^{(t)},x_{te}^{(t)})),
\end{split}
\end{equation}
which means the goal is to find an optimal $\theta$ that can maximize the probability of getting correct $y_{te}$ for all of the given test examples. 

In this case, it is difficult to directly find the optimal $\theta$ since it is a latent variable that is not explicitly defined. A heuristic method is to first assume a desired $\theta$ which can be denoted as $\theta_0$. Under this setting, to enhance the informativeness of the selected ICL demonstrations, we maximize how much the selected demonstrations can change the latent concept from an initial pattern to a desired pattern. This is captured by the difference in log-probabilities, which can be formulated as:
\begin{equation}
\underset{S^{(t)}}{\max}\ [\log P(\theta_{0}|S^{(t)}) - \log P(\theta^{pre}|S^{(t)})].\label{eq1}
\end{equation}
where $\theta^{pre}$ is the pre-trained latent concept while only the task description is used as the prompt without any ICL demonstrations. 

This formulation can be further expanded according to the Bayes rules and simplified as:
\begin{equation}
\begin{split}
P(\theta_{0}|S^{(t)}) &= \frac{P(y_{D}|\theta_{0}, x_{D})P(\theta_{0}|X=x_{D})P(x_{D})}{P(y_{D}|x_{D})P(x_{D})} \\&= \frac{P(y_{D}|\theta_{0},x_{D})P(\theta_{0}|x_{D})}{P(y_{D}|x_{D})},
\end{split}
\end{equation}

\begin{equation}
\begin{split}
P(\theta^{pre}|S^{(t)}) &= \frac{P(y_{D}|\theta^{pre},x_{D})P(\theta^{pre}|x_{D})P(x_{D})}{P(y_{D}|x_{D})P(x_{D})} \\&= \frac{P(y_{D}|\theta^{pre},x_{D})P(\theta^{pre}|x_{D})}{P(y_{D}|x_{D})}.
\end{split}
\end{equation}

The simplified terms are used to substitute $P(\theta_{0}|S^{(t)})$ and $P(\theta^{pre}|S^{(t)})$ in Eq. (\ref{eq1}), which can be written as:
\begin{equation}
\begin{split}
 &\underset{S^{(t)}}{\max}\ \log \frac{P(\theta_{0}|S^{(t)})}{P(\theta^{pre}|S^{(t)})}.\label{eq29}
\\=& \underset{S^{(t)}}{\max}\ \log \frac{P(y_{D}|\theta^{pre},x_{D})P(\theta^{pre}|x_{D})}{P(y_{D}|\theta_{0},x_{D})P(\theta_{0}|x_{D})}
\end{split}
\end{equation}    

Since general-purpose LLM without fine-tuning on the target task is used in this work, the model lacks strong prior preferences over latent concepts for any given inputs. The variation of $P(\theta|x_{D})$ is typically smaller than that of the likelihood term $P(y_{D}|\theta,x_{D})$. In this case, we can assume that $P(\theta|x_{D})$ is approximately uniform across $\theta_{0}$ and $\theta^{pre}$. Under this assumption, the difference in the priors becomes negligible. In this way, it can be simplified as:

\begin{equation}
\underset{x_{D}}{\max}\ [\log P(y_{D}|\theta_{0},x_{D}) - \log P(y_{D}|\theta^{pre},x_{D})].
\end{equation}

An intuitive explanation of this formulation is presented as follows. It measures the difference between the zero-shot predicted results and the expected output of the ICL demonstrations. This indicates how much information about the expected prediction for the given ICL demonstration is not contained in the pre-trained parameters of the LLM. Informative examples are those that cannot be accurately predicted by pre-trained models and, therefore, require ICL.
The output of LLM can be viewed as a sequence formed by multiple digits, where each digit follows its own probability distribution. However, these distributions are interdependent, making it difficult to directly compute the overall probability distribution of the final output. Considering the semantic meaning of the output values, the prediction error rate of ICL demonstrations $Er$ is defined to measure the difference:
\begin{equation}
    Er = \frac{|\hat{y}_{D}-y_{D}|}{y_{D}},
\end{equation}
where $\hat{y_{D}}$ is the zero-shot predicted results of ICL demonstration $x_{D}$:
\begin{equation}
    \hat{y}_{D} \sim P(y|\theta^{pre},x_{D}).
\end{equation}

In the second step of ICL demonstration selection, the prediction error rate $Er$ is calculated for the demonstration selected in the first step and a threshold is set. If the error rate is above the threshold, the demonstration is considered to be informative and the ICL is performed. If the error value is below the threshold, the sample is considered to be insufficiently informative and ICL is not required for the given demonstration.

Specifically, in the first step, our proposed method identifies past traffic patterns that are similar to the current input. Then, it evaluates the informativeness by
checking whether the LLM can already predict the outcome of each selected demonstration in a zero-shot
setting. If it can, the demonstration is considered redundant and excluded. If not, the demonstration is more likely to contain useful information and is kept. These two steps ensures that each selected example is both effective and informative. So our proposed selection method can improve the prediction performance without adding unnecessary length to the prompt.

In the following, we analyze how the second step of ICL demonstration selection influences the prediction error according to Eq. \eqref{eq22} and Eq. \eqref{eq23}. If a selected demonstration has a low zero-shot prediction error, two cases may arise. The first case is that the test sample is similar to the selected demonstration. In this case, a low prediction error of the selected demonstration means that the LLM’s prediction error for the test sample, $\epsilon_{LLM}$, is also low. The second case is that although the selected demonstration has a low zero-shot prediction error, the test sample is not similar enough to the selected demonstration. In this case, the kernel product of the test sample input and selected demonstration input is small. As a result, in both cases, using the ICL demonstrations selected in the first step may not significantly reduce prediction error. Therefore, in this step, we filter out ICL demonstrations that do not provide additional performance improvement, ensuring that only ICL demonstrations that reduce prediction error are used.

\subsection{Computational complexity of proposed method}

In this subsection, we discuss the computational complexity of the proposed demonstration selection method. For the first step, the calculations are performed during the test time. This step involves traversing all available ICL demonstrations and computing the metrics $e_1$ and $e_2$ for each demonstration. Consequently, the computational complexity of this step is proportional to the number of available ICL demonstrations, denoted as $n_d$, and can be expressed as $O(n_d)$.

For the second step, part of the calculations can be performed offline before the test phase. For instance, the calculation of the predicted error rate $Er$ is independent of the specific test case and can be completed in advance. In this case, during the test time, the only calculation required is to compare the predicted error rates of selected ICL demonstrations with the threshold value, which is computationally negligible. Therefore, the overall computational complexity of the proposed method remains $O(n_d)$ for selecting a single demonstration. For multiple selected demonstrations, the overall computational complexity is $O(n_dn_s)$, where $n_s$ denotes the number of selected demonstrations.

In the following, a comparative analysis of computational complexity with other selection methods is provided. ICL demonstration selection methods generally fall into two categories: gradient-based and retrieval-based selection. Gradient-based methods usually require multiple model evaluations, resulting in a significantly higher complexity, $O(N_dm_{llm})$, where $m_{llm}$ denotes the size of the LLM. So the Gradient-based selection methods have a much higher complexity than our proposed method. Retrieval-based methods select the nearest demonstration based on a similarity metric with a complexity of $O(n_dn_s)$.

The first step of our method can be regarded as a retrieval-based selection step. The difference is that our approach includes the second step with an additional pre-processing step to calculate the predicted error rates of demonstrations. Since this part of the calculation can be performed offline before the test phase, the total computation of the proposed selection approach during the test phase is comparable with the standard retrieval-based methods but increases the prediction performance.

In conclusion, our method achieves a balance between computational efficiency and performance by maintaining linear complexity while enhancing the prediction performance.

\subsection{Batch calibration}
Due to the current limitations of LLMs, the predictive results using a single ICL demonstration may suffer from instability. However, for multiple ICL demonstrations, the performance of ICL is usually sensitive to the permutation of demonstrations \cite{wu2023self}. On the other hand, the permutation problem of ICL demonstrations is complex, and finding the optimal permutation directly is very difficult. 

In this work, a batch calibration method is adopted to mitigate the influence of permutation with multiple ICL demonstrations. Specifically, other than performing M-shot prediction with all the selected ICL demonstrations at one time, the proposed method performs one-shot prediction separately with different ICL demonstrations. Then the outputs of one-shot predictions are averaged as the final result. This process can be formulated as:
\begin{equation}
    \hat{y}_{te} =\frac{1}{M}\sum_{i_1}^{i_m}\hat{y}_{te}^{(i_m)},
\end{equation}
where:
\begin{equation}
    \hat{y}_{te}^{(i_m)} \sim P(y|\theta^{pre},(\Gamma_{tr}^{(i_m)}, Z_{tr}^{(i_m)},y_{tr}^{(i_m)}),x_{te}).
\end{equation}
$\hat{y}_{te}^{(i_m)}$ denotes one-shot prediction results with the $i_m^{th}$ ICL demonstration. $i_m$ is the index of the $m^{th}$ selected ICL demonstration. With this method, the robustness and accuracy of LLM-based prediction can be increased.

\section{Numerical Results}

\subsection{Experimental Settings and Evaluation Metrics}
In this work, the real-world 5G dataset in \cite{raca2020beyond} is used. The dataset includes two kinds of data, static data and driving data, and it also includes three different activities, namely downloading files, watching videos on Netflix, and watching videos on Amazon. The dataset also considers the co-existence of multiple technologies, including 5G, LTE, and HSPA+, which adds extra difficulties to the prediction task and makes it more comparable to real-world situations.

The dataset includes four types of contextual information, the throughput, the channel metrics, the neighboring cell metrics, and other context metrics. There are 25 contextual features in total. In addition to the historical downlink throughput feature, we selected four contextual features for the downlink traffic prediction task: uplink throughput, the reference signal received power (RSRP) of the currently connected cell, the RSRP of the neighboring cell, and the network mode. The feature selection was based on the feature ranking results in \cite{mei2022realtime}, where the authors calculated the feature importance scores using the same dataset used in this study. Based on these scores, we selected the four most influential contextual features to enhance prediction performance.

Unlike traditional ML-based traffic prediction methods, this work does not apply other pre-processing steps such as normalization or standardization. The primary reason is to preserve the semantic integrity of numerical values. Since LLMs are pre-trained on diverse textual and numerical data, they can interpret raw numerical values without requiring rescaling. Similarly, pre-processing steps such as one-hot encoding is not used for categorical features because LLMs can directly process natural language-based inputs.

During the experiments, we equally divided the whole dataset into two parts. The first part is used for ICL demonstration selection and the second part is used for testing. The historical time window $H$ is set as 5. 

\begin{figure*}[t]
    \centering
    \subfigure[Driving, and downloading files.]{
        \includegraphics[width=8cm,height=5.5cm]{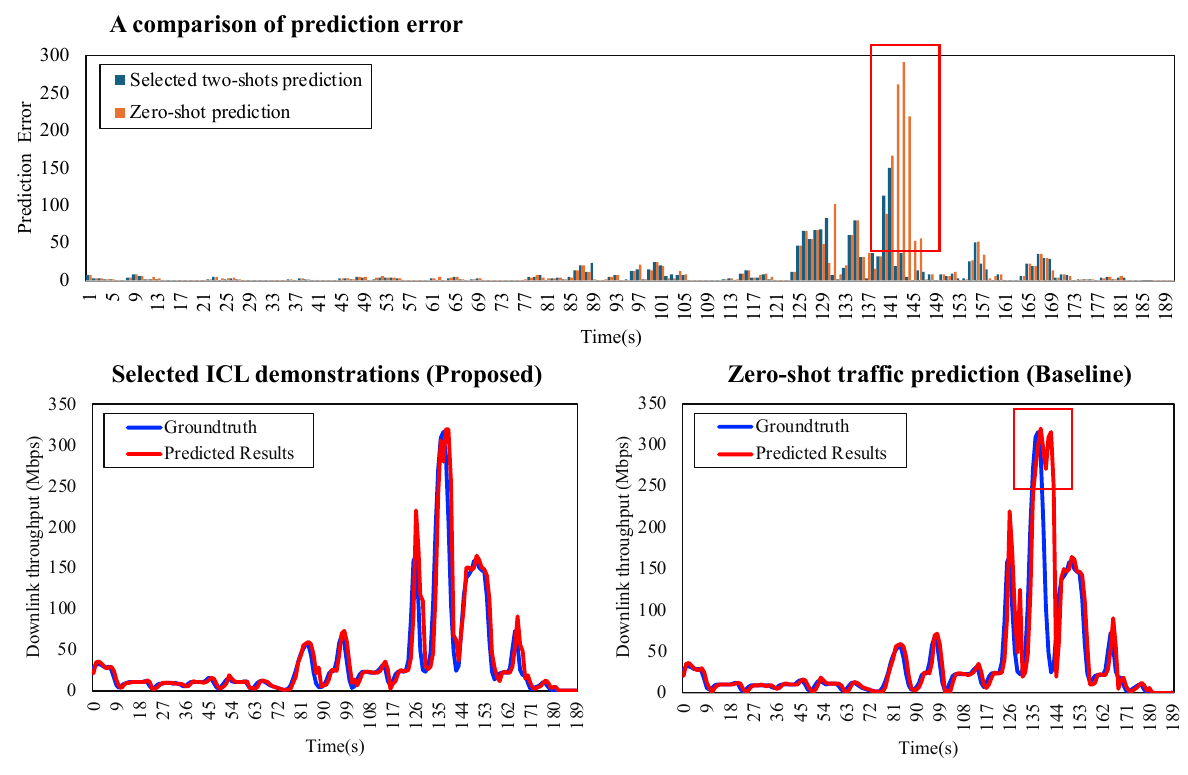}
    }
    \subfigure[Static, and downloading files.]{
	\includegraphics[width=8cm,height=5.5cm]{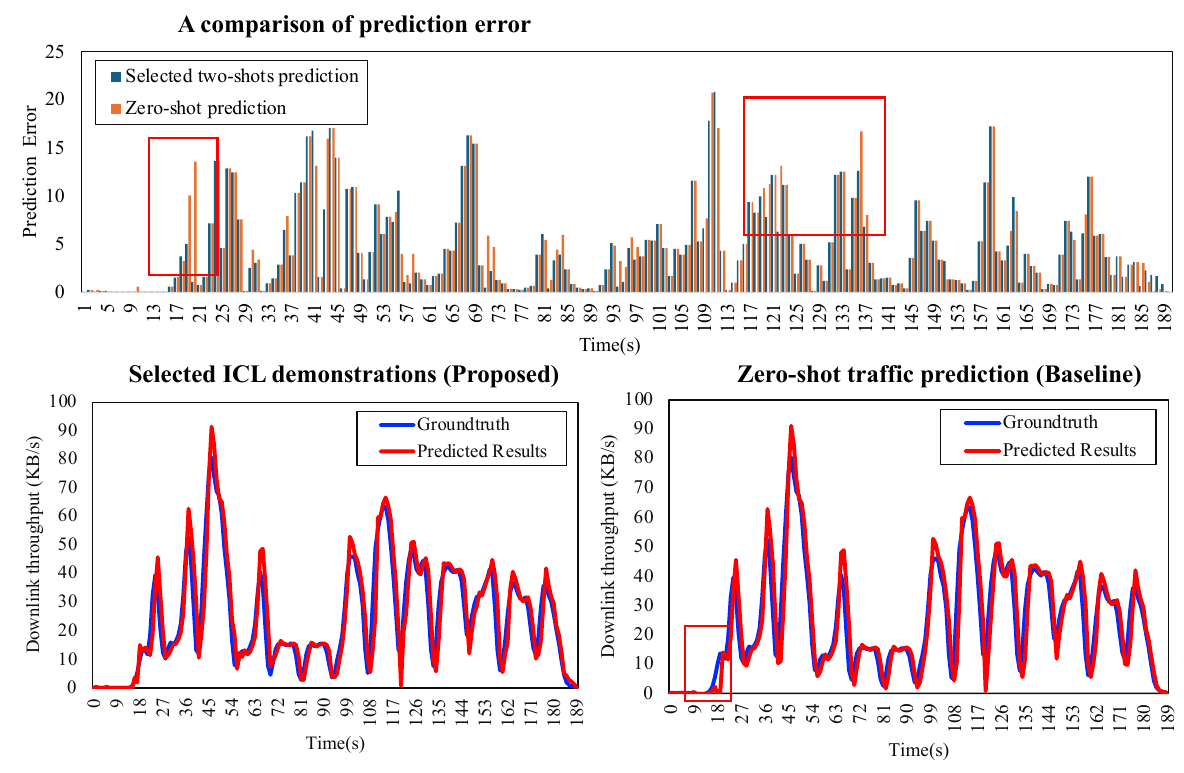}
    }
    \subfigure[Driving, and watching Amazon videos.]{
	\includegraphics[width=8cm,height=5.5cm]{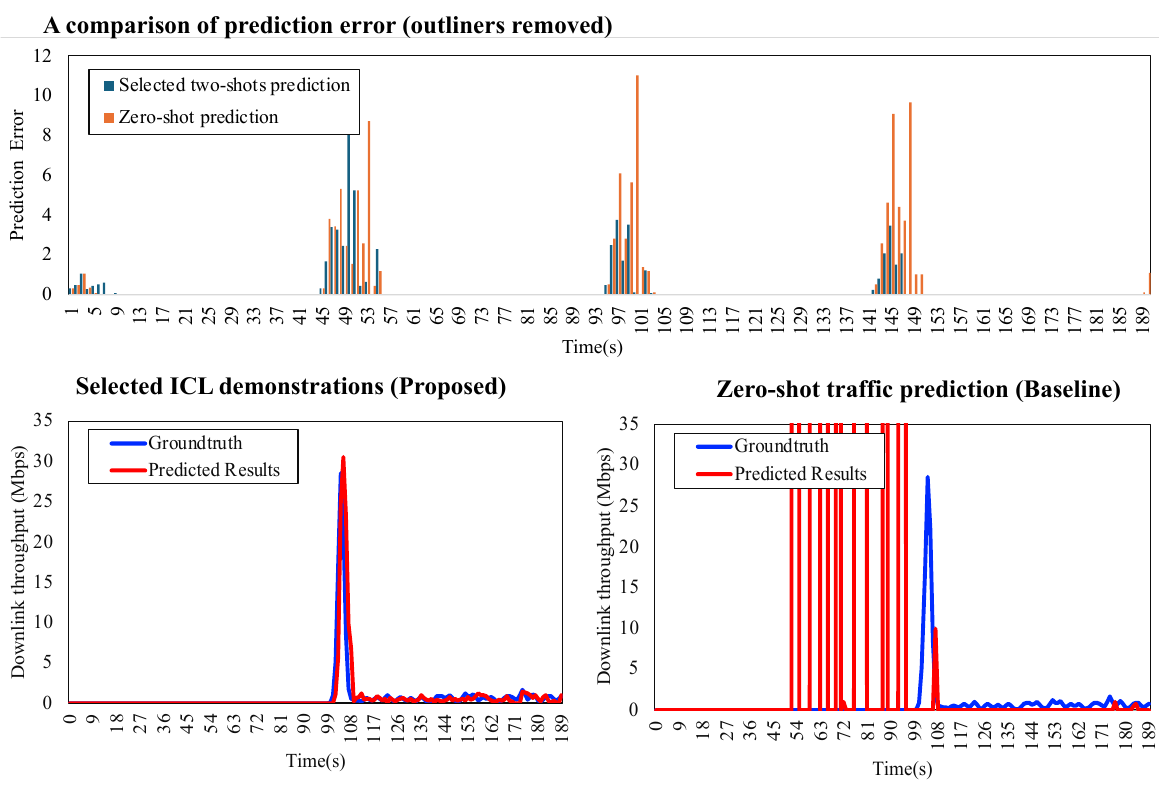}
    }
    \subfigure[Driving, and watching Netflix videos.]{
	\includegraphics[width=8cm,height=5.5cm]{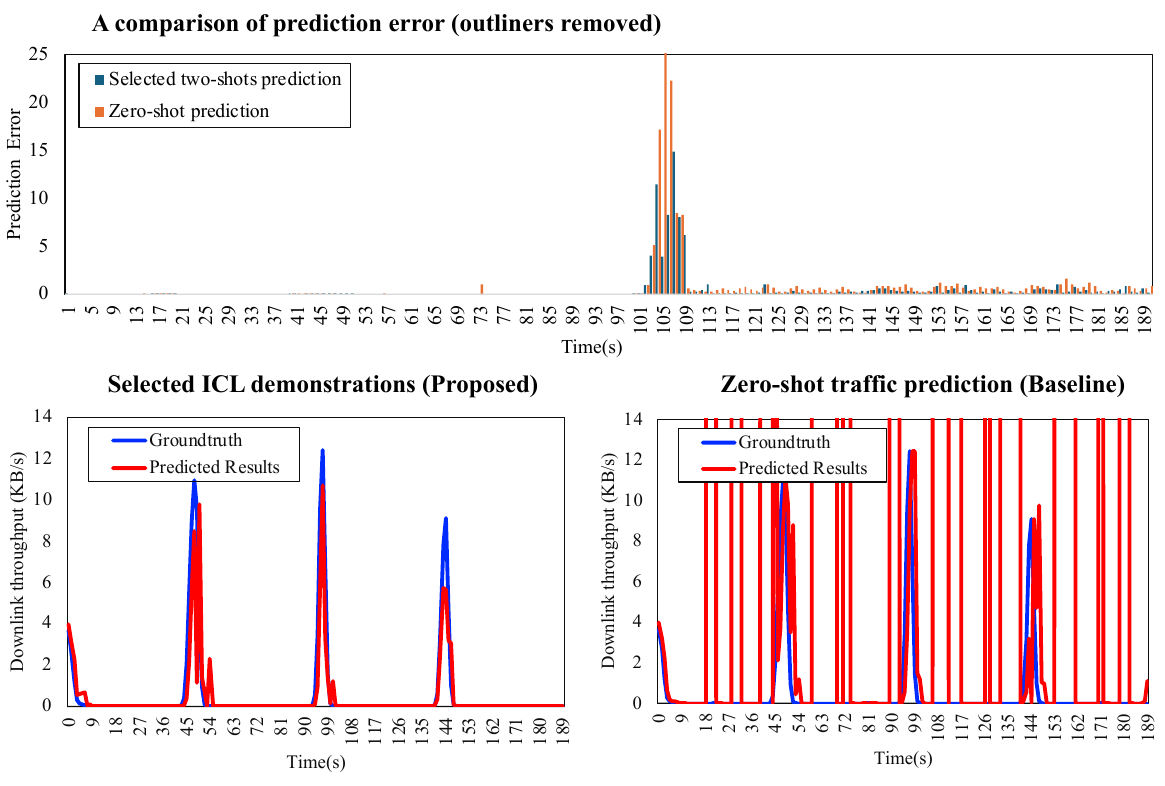}
    }
    \caption{A comparison of traffic prediction with the proposed selected ICL demonstrations and zero-shot traffic prediction in four scenarios using PHI-3-medium.}\label{fig3-new}

\end{figure*}

Three indicators are used to evaluate the prediction performance, the mean absolute error (MAE), the root mean squared error (RMSE), and the $R^2$-Score. MAE measures the average of the absolute errors between predicted and actual values. It can be calculated as:
\begin{equation}
    MAE = \frac{1}{T}\sum_{t}|y^{(t)}_{te}-\hat{y}^{(t)}_{te}|.
\end{equation}

RMSE measures the standard deviation of the predicted errors. It can be calculated as:
\begin{equation}
    RMSE = \sqrt{\frac{1}{T}\sum_{t}(y^{(t)}_{te}-\hat{y}^{(t)}_{te})^2}.
\end{equation}

$R^2$-Score is a commonly used metric to evaluate how well the model predicts the outcome of the dependent variable. It can be calculated as:
\begin{equation}
    R^2-Score = 1-\frac{\sum_{t}(y^{(t)}_{te}-\hat{y}^{(t)}_{te})^2/T}{\sum_{t}(y^{(t)}_{te}-\overline{y_{te}})^2/T}.
\end{equation}

In this work, we use PHI-3 model as the pre-trained LLM for prediction \cite{abdin2024phi}. The PHI-3 model is a fully open-source language model that ranges in size from 3.8 billion to 14 billion parameters depending on the version. If not otherwise specified, we use PHI-3-medium with 14 billion model parameters as the testing model. In the next section, the numerical results are provided to demonstrate the effectiveness of our proposed method.

\begin{table*}[]
\centering
\caption{MAE, RMSE and $R^2$-Score of mobile traffic prediction under different scenarios.}
\label{table1}
\resizebox{2\columnwidth}{!}{%
\begin{tabular}{|l|lll|lll|lll|lll|}
\hline
 &
   &
   &
   &
   &
   &
   &
   &
   &
   &
   &
   &
   \\
\multicolumn{1}{|c|}{} &
  \multicolumn{3}{c|}{Downloading files - driving} &
  \multicolumn{3}{c|}{Downloading files - static} &
  \multicolumn{3}{c|}{Watching Amazon videos - driving} &
  \multicolumn{3}{c|}{Watching Netflix videos - driving} \\
 &
   &
   &
   &
   &
   &
   &
   &
   &
   &
   &
   &
   \\ \hline
 &
  \multicolumn{1}{l|}{} &
  \multicolumn{1}{l|}{} &
   &
  \multicolumn{1}{l|}{} &
  \multicolumn{1}{l|}{} &
   &
  \multicolumn{1}{l|}{} &
  \multicolumn{1}{l|}{} &
   &
  \multicolumn{1}{l|}{} &
  \multicolumn{1}{l|}{} &
   \\
\multicolumn{1}{|c|}{} &
  \multicolumn{1}{c|}{\begin{tabular}[c]{@{}c@{}}Selected \\ two-demo \\ prediction \\ (Proposed)\end{tabular}} &
  \multicolumn{1}{c|}{\begin{tabular}[c]{@{}c@{}}Zero-shot \\ prediction\end{tabular}} &
  \multicolumn{1}{c|}{\begin{tabular}[c]{@{}c@{}}Weighted \\ moving \\ average\end{tabular}} &
  \multicolumn{1}{c|}{\begin{tabular}[c]{@{}c@{}}Selected \\ two-demo \\ prediction \\ (Proposed)\end{tabular}} &
  \multicolumn{1}{c|}{\begin{tabular}[c]{@{}c@{}}Zero-shot\\  prediction\end{tabular}} &
  \multicolumn{1}{c|}{\begin{tabular}[c]{@{}c@{}}Weighted \\ moving \\ average\end{tabular}} &
  \multicolumn{1}{c|}{\begin{tabular}[c]{@{}c@{}}Selected \\ two-demo \\ prediction \\ (Proposed)\end{tabular}} &
  \multicolumn{1}{c|}{\begin{tabular}[c]{@{}c@{}}Zero-shot\\  prediction\end{tabular}} &
  \multicolumn{1}{c|}{\begin{tabular}[c]{@{}c@{}}Weighted \\ moving \\ average\end{tabular}} &
  \multicolumn{1}{c|}{\begin{tabular}[c]{@{}c@{}}Selected \\ two-demo \\ prediction \\ (Proposed)\end{tabular}} &
  \multicolumn{1}{c|}{\begin{tabular}[c]{@{}c@{}}Zero-shot\\  prediction\end{tabular}} &
  \multicolumn{1}{c|}{\begin{tabular}[c]{@{}c@{}}Weighted \\ moving \\ average\end{tabular}} \\
 &
  \multicolumn{1}{l|}{} &
  \multicolumn{1}{l|}{} &
   &
  \multicolumn{1}{l|}{} &
  \multicolumn{1}{l|}{} &
   &
  \multicolumn{1}{l|}{} &
  \multicolumn{1}{l|}{} &
   &
  \multicolumn{1}{l|}{} &
  \multicolumn{1}{l|}{} &
   \\ \hline
 &
  \multicolumn{1}{l|}{} &
  \multicolumn{1}{l|}{} &
   &
  \multicolumn{1}{l|}{} &
  \multicolumn{1}{l|}{} &
   &
  \multicolumn{1}{l|}{} &
  \multicolumn{1}{l|}{} &
   &
  \multicolumn{1}{l|}{} &
  \multicolumn{1}{l|}{} &
   \\
\multicolumn{1}{|c|}{MAE} &
  \multicolumn{1}{c|}{\textbf{9.4913}} &
  \multicolumn{1}{c|}{13.816} &
  \multicolumn{1}{c|}{16.052} &
  \multicolumn{1}{c|}{\textbf{4.4605}} &
  \multicolumn{1}{c|}{4.7327} &
  \multicolumn{1}{c|}{8.0972} &
  \multicolumn{1}{c|}{\textbf{0.42907}} &
  \multicolumn{1}{c|}{0.73560} &
  \multicolumn{1}{c|}{0.58080} &
  \multicolumn{1}{c|}{\textbf{0.29391}} &
  \multicolumn{1}{c|}{0.86242} &
  \multicolumn{1}{c|}{0.58674} \\
 &
  \multicolumn{1}{l|}{} &
  \multicolumn{1}{l|}{} &
   &
  \multicolumn{1}{l|}{} &
  \multicolumn{1}{l|}{} &
   &
  \multicolumn{1}{l|}{} &
  \multicolumn{1}{l|}{} &
   &
  \multicolumn{1}{l|}{} &
  \multicolumn{1}{l|}{} &
   \\ \hline
 &
  \multicolumn{1}{l|}{} &
  \multicolumn{1}{l|}{} &
   &
  \multicolumn{1}{l|}{} &
  \multicolumn{1}{l|}{} &
   &
  \multicolumn{1}{l|}{} &
  \multicolumn{1}{l|}{} &
   &
  \multicolumn{1}{l|}{} &
  \multicolumn{1}{l|}{} &
   \\
\multicolumn{1}{|c|}{RMSE} &
  \multicolumn{1}{c|}{\textbf{21.827}} &
  \multicolumn{1}{c|}{39.584} &
  \multicolumn{1}{c|}{32.848} &
  \multicolumn{1}{c|}{\textbf{6.3854}} &
  \multicolumn{1}{c|}{6.6346} &
  \multicolumn{1}{c|}{11.125} &
  \multicolumn{1}{c|}{\textbf{1.7275}} &
  \multicolumn{1}{c|}{3.0858} &
  \multicolumn{1}{c|}{2.5201} &
  \multicolumn{1}{c|}{\textbf{1.0267}} &
  \multicolumn{1}{c|}{3.2867} &
  \multicolumn{1}{c|}{1.7135} \\
 &
  \multicolumn{1}{l|}{} &
  \multicolumn{1}{l|}{} &
   &
  \multicolumn{1}{l|}{} &
  \multicolumn{1}{l|}{} &
   &
  \multicolumn{1}{l|}{} &
  \multicolumn{1}{l|}{} &
   &
  \multicolumn{1}{l|}{} &
  \multicolumn{1}{l|}{} &
   \\ \hline
 &
  \multicolumn{1}{l|}{} &
  \multicolumn{1}{l|}{} &
   &
  \multicolumn{1}{l|}{} &
  \multicolumn{1}{l|}{} &
   &
  \multicolumn{1}{l|}{} &
  \multicolumn{1}{l|}{} &
   &
  \multicolumn{1}{l|}{} &
  \multicolumn{1}{l|}{} &
   \\
\multicolumn{1}{|c|}{$R^2$-Score} &
  \multicolumn{1}{c|}{\textbf{0.84743}} &
  \multicolumn{1}{c|}{0.49822} &
  \multicolumn{1}{c|}{0.77634} &
  \multicolumn{1}{c|}{\textbf{0.87349}} &
  \multicolumn{1}{c|}{0.87434} &
  \multicolumn{1}{c|}{0.61598} &
  \multicolumn{1}{c|}{\textbf{0.65831}} &
  \multicolumn{1}{c|}{-0.0902} &
  \multicolumn{1}{c|}{0.27286} &
  \multicolumn{1}{c|}{\textbf{0.74821}} &
  \multicolumn{1}{c|}{-1.5804} &
  \multicolumn{1}{c|}{0.29866} \\
 &
  \multicolumn{1}{l|}{} &
  \multicolumn{1}{l|}{} &
   &
  \multicolumn{1}{l|}{} &
  \multicolumn{1}{l|}{} &
   &
  \multicolumn{1}{l|}{} &
  \multicolumn{1}{l|}{} &
   &
  \multicolumn{1}{l|}{} &
  \multicolumn{1}{l|}{} &
   \\ \hline
\end{tabular}%
}
\end{table*}

\subsection{Performance Analysis of ICL}



In this subsection, we analyze how ICL influences the predictive results. We first perform traffic prediction in four different application scenarios, and compare the predictive results obtained using our proposed ICL method with the zero-shot prediction results. During the experiments, we select two ICL demonstrations since more demonstrations will lead to a higher computation cost and according to the numerical result, two demonstrations already can lead to a better performance and eliminate noise. It should be noted that the choice of the number of ICL demonstrations is also highly dependent on the size of the dataset and cannot be generalized. Using the batch calibration method makes it easier to manage and adjust the number of demonstrations without restructuring the prompt. In practice, we observe that adding more demonstrations does not always lead to better performance. When too many demonstrations are used, additional examples may introduce noise or redundancy. To decide the optimal number of ICL demonstrations, we can first consider the inference time requirements and available computational resources. Then, we can observe how model performance changes as the number of demonstrations increases. A simple approach is to gradually increase the number of demonstrations and monitor the predictions. If the results become more stable and accurate at first but then stop improving or even begin to degrade, we can choose the number just before that turning point.

Fig. \ref{fig3-new} shows the traffic prediction results using PHI-3-medium with selected ICL demonstrations, and zero-shot prediction results. The term zero-shot refers to predictions made without incorporating any ICL demonstrations in the prompt and serves as a baseline. To better highlight the differences between the predicted and ground-truth values over time, we have also included the prediction error bars for both the proposed method and the zero-shot baseline for comparison. The prediction error bars of our proposed method is shown in blue color while the zero-shot baseline is shown in orange color.

As illustrated in Fig. \ref{fig3-new}, our proposed ICL method improves both the accuracy and robustness of predictions across different application scenarios. In the first two scenarios, the performance difference between predictions with and without ICL is relatively small. But it can still be observed that the performance with selected ICL demonstrations is slightly better than zero-shot predictions and some noticeable differences are marked with red boxes.

In contrast, the third and fourth scenarios exhibit significant performance degradation under zero-shot prediction. In particular, some predicted values are much higher than the typical traffic bitrate range. These values are generally due to the inability of the LLM to correctly output the end mark, and they are considered hallucinations. The use of selected ICL demonstrations effectively mitigates these hallucination issues, leading to more accurate and stable predictions.

To further analyze prediction performance, in the visualization of the prediction errors, we remove these unstable outlier points. The remaining data shows that predictions made with selected ICL demonstrations consistently still have lower errors compared to zero-shot predictions. This further demonstrate the effectiveness of ICL in enhancing both stability and accuracy.

Table \ref{table1} shows the MAE, RMSE, and $R^2$-Score of the traffic predictions under different application scenarios. The $R^2$-Score ranges from $-\infty$ to 1, where a score closer to 1 reflects better predictive performance. The predictive results with the weighted moving average model \cite{kapgate2014weighted} are also added for comparison. To enable a better comparison, the hallucinations of zero-shot prediction have been manually removed. As can be observed from the table, our proposed ICL-based traffic prediction method achieves lower MAE, lower RMSE, and higher $R^2$-Scores compared with the other two baselines. The $R^2$-Scores of the ICL-based predictions under different scenarios show that the proposed algorithms perform better in scenarios with gradual changes in traffic flow and perform worse in scenarios with abrupt changes in traffic flow. However, in all the application scenarios, the $R^2$-Scores are above 0.65, indicating that using ICL, the pre-trained LLM can successfully predict the mobile traffic in different scenarios.

\begin{figure*}[t]
    \centering
    \subfigure[A comparison of the proposed similarity measurement distance-based similarity measurement under varying demonstrations.]{
        \includegraphics[width=12cm]{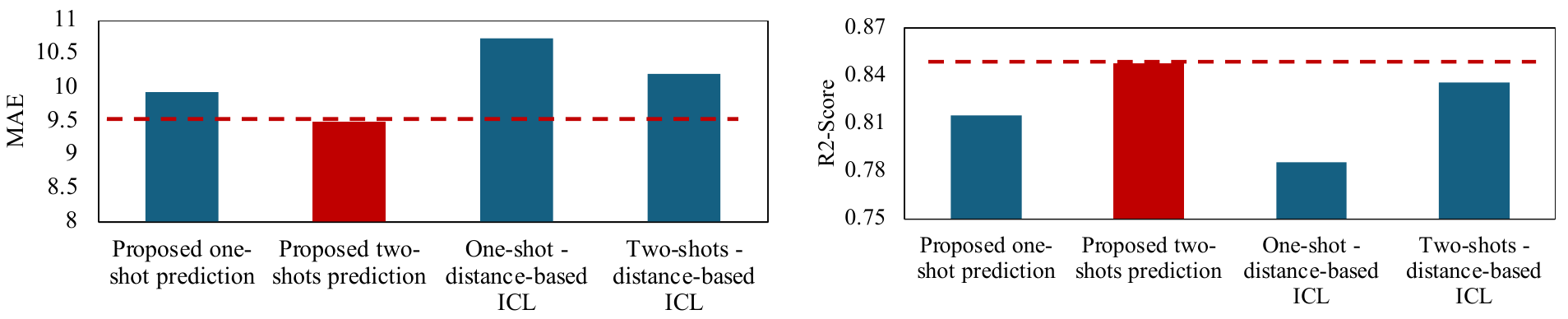}
    }
    \subfigure[A comparison of the proposed method and other traffic prediction methods.]{
	\includegraphics[width=12cm]{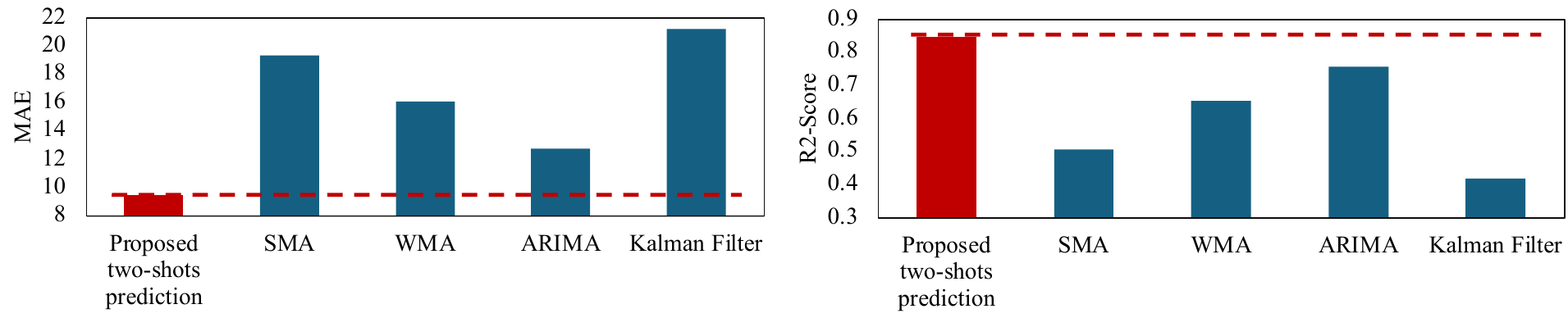}
    }
    \subfigure[An ablation study on effectiveness and informativeness rules.]{
	\includegraphics[width=12cm]{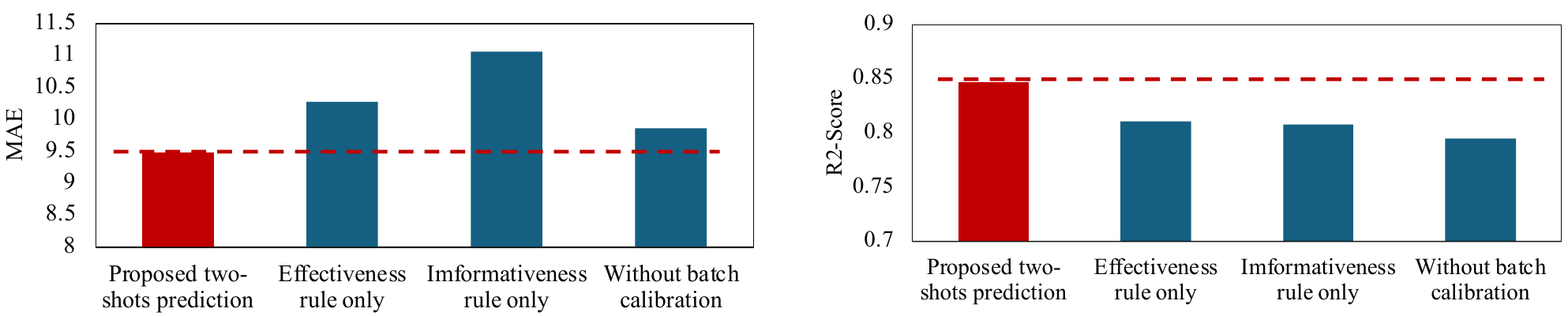}
    }
    \caption{MAE and $R^2$-Score of mobile traffic prediction under different ICL settings.}
    \label{figabl}
\end{figure*}

Fig. \ref{figabl} further compares the MAE and $R^2$-Scores of mobile traffic prediction under different ICL settings in the first scenario.

Fig. \ref{figabl}(a) assesses the impact of our proposed similarity evaluation method by replacing it with a standard metric based on time series distance measures, keeping the rest of the pipeline identical. According to the simulation results, we show that our similarity evaluation method is more reasonable than the standard distance metric.

Fig. \ref{figabl}(b) shows a comparison of the proposed method with three other traffic prediction methods, the simple moving average (SMA) method, the weighted moving average (WMA) method, the autoregressive integrated moving average (ARIMA) method and the Kalman Filter-based method. Our proposed method is highlighted in red. According to the results, our method consistently achieves lower MAE and higher $R^2$ than the other four baseline methods.

Fig. \ref{figabl}(c) shows a comparison of the MAE and the $R^2$-score for three ablation settings: effectiveness-based selection only, informativeness-based selection only, and the full two-step selection mechanism without batch calibration. As shown in the figure, all three ablated variants result in degraded performance compared to our proposed full two-step selection method. When using only the effectiveness rule, the selected demonstrations may lack sufficient information and lead to redundancy in the prompt. Conversely, when using only the informativeness rule, the selected demonstrations may be less relevant to the test instance and have limited effects on the prediction accuracy. These results highlight the complementary roles of both effectiveness and informativeness in our selection strategy and demonstrate the necessity of each step in achieving optimal performance.

\begin{table*}[]
\centering
\caption{The ICL ratio, MAE, RMSE and $R^2$-Score under different predicted error rate threshold settings.}
\label{table3}
\resizebox{1.5\columnwidth}{!}{%
\begin{tabular}{|l|l|l|l|l|l|l|l|}
\hline
 &
   &
   &
   &
   &
   &
   &
   \\
\multicolumn{1}{|c|}{} &
  \multicolumn{1}{c|}{\begin{tabular}[c]{@{}c@{}}Zero-shot \\ prediction\end{tabular}} &
  \multicolumn{1}{c|}{\begin{tabular}[c]{@{}c@{}}Threshold \\ = 0.8\end{tabular}} &
  \multicolumn{1}{c|}{\begin{tabular}[c]{@{}c@{}}Threshold \\ = 0.5\end{tabular}} &
  \multicolumn{1}{c|}{\begin{tabular}[c]{@{}c@{}}Threshold \\ = 0.35\end{tabular}} &
  \multicolumn{1}{c|}{\begin{tabular}[c]{@{}c@{}}Threshold \\ = 0.15\end{tabular}} &
  \multicolumn{1}{c|}{\begin{tabular}[c]{@{}c@{}}Threshold \\ = 0.05\end{tabular}} &
  \multicolumn{1}{c|}{Full ICL} \\
 &
   &
   &
   &
   &
   &
   &
   \\ \hline
 &
   &
   &
   &
   &
   &
   &
   \\
ICL ratio &
  \multicolumn{1}{c|}{0.00\%} &
  \multicolumn{1}{c|}{41.20\%} &
  \multicolumn{1}{c|}{47.60\%} &
  \multicolumn{1}{c|}{47.60\%} &
  \multicolumn{1}{c|}{50.26\%} &
  \multicolumn{1}{c|}{82.01\%} &
  \multicolumn{1}{c|}{100.00\%} \\
 &
   &
   &
   &
   &
   &
   &
   \\ \hline
 &
   &
   &
   &
   &
   &
   &
   \\
\multicolumn{1}{|c|}{$R^2$-Score} &
  \multicolumn{1}{c|}{0.49822} &
  \multicolumn{1}{c|}{0.75045} &
  \multicolumn{1}{c|}{0.83880} &
  \multicolumn{1}{c|}{0.84744} &
  \multicolumn{1}{c|}{\textbf{0.84873}} &
  \multicolumn{1}{c|}{0.80990} &
  \multicolumn{1}{c|}{0.81111} \\
 &
   &
   &
   &
   &
   &
   &
   \\ \hline
 &
   &
   &
   &
   &
   &
   &
   \\
\multicolumn{1}{|c|}{MAE} &
  \multicolumn{1}{c|}{13.816} &
  \multicolumn{1}{c|}{11.301} &
  \multicolumn{1}{c|}{9.6202} &
  \multicolumn{1}{c|}{9.4913} &
  \multicolumn{1}{c|}{\textbf{9.4263}} &
  \multicolumn{1}{c|}{10.348} &
  \multicolumn{1}{c|}{10.271} \\
 &
   &
   &
   &
   &
   &
   &
   \\ \hline
 &
   &
   &
   &
   &
   &
   &
   \\
\multicolumn{1}{|c|}{RMSE} &
  \multicolumn{1}{c|}{39.585} &
  \multicolumn{1}{c|}{27.916} &
  \multicolumn{1}{c|}{22.437} &
  \multicolumn{1}{c|}{21.827} &
  \multicolumn{1}{c|}{\textbf{21.735}} &
  \multicolumn{1}{c|}{24.365} &
  \multicolumn{1}{c|}{24.287} \\
 &
   &
   &
   &
   &
   &
   &
   \\ \hline
\end{tabular}%
}
\end{table*}

\subsection{Performance Analysis under Different Predicted Error Rate Thresholds}
In the proposed two-step efficient demonstration selection method, the predicted error rate is used to evaluate the informativeness of selected ICL demonstrations. In this section, we discuss how different predicted error rate thresholds can influence the prediction performance. With these numerical results, we demonstrate the necessity and validity of evaluating the informativeness of ICL demonstrations.

\begin{figure}[h]
\centering
\includegraphics[width=8cm]{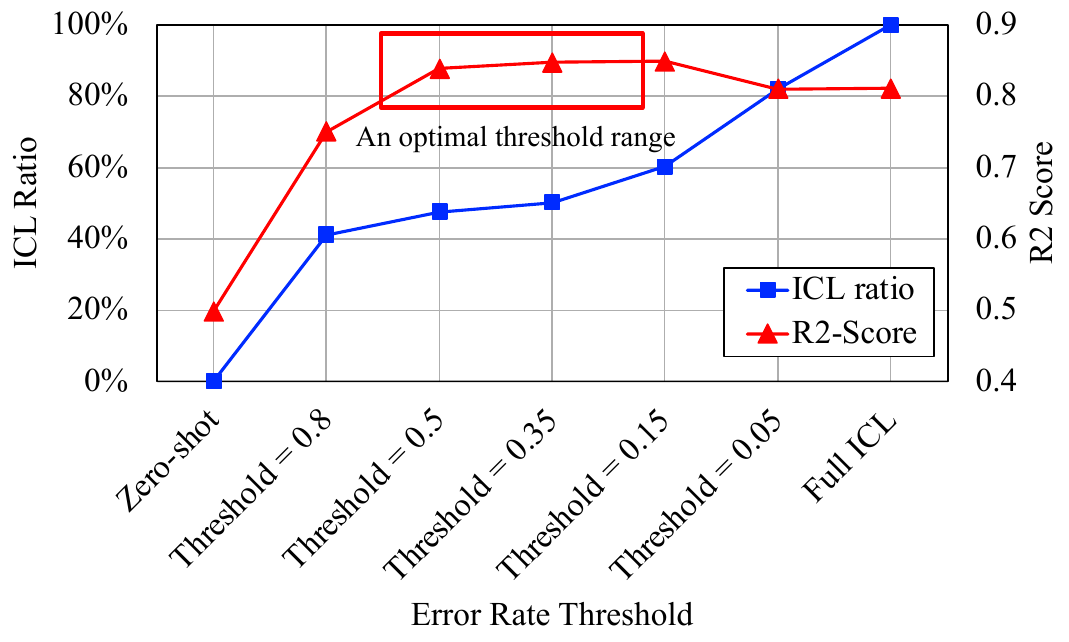}
\caption{The ICL ratio and the $R^2$-Score under different predicted error rate threshold settings.}
\label{fig4}
\end{figure}

Fig. \ref{fig4} and Table \ref{table3} show how the setting of the predicted error rate threshold affects the prediction performance. Fig. \ref{fig4} shows the trend in the ICL ratio and the $R^2$-Scores for different error rate threshold settings. The ICL ratio means the ratio of test instances where ICL is used for prediction. In Table \ref{table3}, specific values of ICL ratio, MAE, RMSE, and $R^2$-Scores are given for different prediction error rate threshold settings. As can be observed from the figure and the table, if the threshold is set lower, the ICL ratio will be higher, which means that ICL will be performed more frequently with the test instances. Meanwhile, the $R^2$-Score first increases when the threshold is greater than 0.5, becomes relatively stable when the threshold is greater than 0.15, and finally decreases slightly when the threshold is less than 0.15. This suggests that not all test instances require ICL. In some test instances, the ICL demonstrations will provide useless information and interfere with predictive results. According to the numerical results, setting the threshold between 0.5 and 0.15 is an ideal range to optimize the prediction accuracy and computational cost of ICL.

\subsection{Performance Analysis of LLM-based Prediction and LSTM-based Prediction}

\begin{table*}[]
\centering
\caption{MAE, RMSE and $R^2$-Score of mobile traffic prediction of LLM-based method and LSTM-based method.}
\label{table6}
\resizebox{1.8\columnwidth}{!}{%
\begin{tabular}{|l|ll|ll|ll|ll|}
\hline
 &
   &
   &
   &
   &
   &
   &
   &
   \\
\multicolumn{1}{|c|}{} &
  \multicolumn{2}{c|}{\begin{tabular}[c]{@{}c@{}}Downloading files\\  - driving\end{tabular}} &
  \multicolumn{2}{c|}{\begin{tabular}[c]{@{}c@{}}Downloading files\\  - static\end{tabular}} &
  \multicolumn{2}{c|}{\begin{tabular}[c]{@{}c@{}}Watching Amazon videos\\  - driving\end{tabular}} &
  \multicolumn{2}{c|}{\begin{tabular}[c]{@{}c@{}}Watching Netflix videos\\  - driving\end{tabular}} \\
 &
   &
   &
   &
   &
   &
   &
   &
   \\ \hline
 &
  \multicolumn{1}{l|}{} &
   &
  \multicolumn{1}{l|}{} &
   &
  \multicolumn{1}{l|}{} &
   &
  \multicolumn{1}{l|}{} &
   \\
\multicolumn{1}{|c|}{} &
  \multicolumn{1}{c|}{\begin{tabular}[c]{@{}c@{}}LLM-based \\ prediction\end{tabular}} &
  \multicolumn{1}{c|}{\begin{tabular}[c]{@{}c@{}}LSTM-based \\ prediction\end{tabular}} &
  \multicolumn{1}{c|}{\begin{tabular}[c]{@{}c@{}}LLM-based\\ prediction\end{tabular}} &
  \multicolumn{1}{c|}{\begin{tabular}[c]{@{}c@{}}LSTM-based\\  prediction\end{tabular}} &
  \multicolumn{1}{c|}{\begin{tabular}[c]{@{}c@{}}LLM-based\\ prediction\end{tabular}} &
  \multicolumn{1}{c|}{\begin{tabular}[c]{@{}c@{}}LSTM-based\\ prediction\end{tabular}} &
  \multicolumn{1}{c|}{\begin{tabular}[c]{@{}c@{}}LLM-based\\ prediction\end{tabular}} &
  \multicolumn{1}{c|}{\begin{tabular}[c]{@{}c@{}}LSTM-based\\  prediction\end{tabular}} \\
 &
  \multicolumn{1}{l|}{} &
   &
  \multicolumn{1}{l|}{} &
   &
  \multicolumn{1}{l|}{} &
   &
  \multicolumn{1}{l|}{} &
   \\ \hline
 &
  \multicolumn{1}{l|}{} &
   &
  \multicolumn{1}{l|}{} &
   &
  \multicolumn{1}{l|}{} &
   &
  \multicolumn{1}{l|}{} &
   \\
\multicolumn{1}{|c|}{MAE} &
  \multicolumn{1}{c|}{9.4913} &
  \multicolumn{1}{c|}{\textbf{7.3187}} &
  \multicolumn{1}{c|}{4.4605} &
  \multicolumn{1}{c|}{\textbf{2.8560}} &
  \multicolumn{1}{c|}{0.42907} &
  \multicolumn{1}{c|}{\textbf{0.22540}} &
  \multicolumn{1}{c|}{\textbf{0.29391}} &
  \multicolumn{1}{c|}{0.30701} \\
 &
  \multicolumn{1}{l|}{} &
   &
  \multicolumn{1}{l|}{} &
   &
  \multicolumn{1}{l|}{} &
   &
  \multicolumn{1}{l|}{} &
   \\ \hline
 &
  \multicolumn{1}{l|}{\textbf{}} &
  \textbf{} &
  \multicolumn{1}{l|}{\textbf{}} &
  \textbf{} &
  \multicolumn{1}{l|}{\textbf{}} &
  \textbf{} &
  \multicolumn{1}{l|}{\textbf{}} &
  \textbf{} \\
\multicolumn{1}{|c|}{RMSE} &
  \multicolumn{1}{c|}{21.827} &
  \multicolumn{1}{c|}{\textbf{17.469}} &
  \multicolumn{1}{c|}{6.3854} &
  \multicolumn{1}{c|}{\textbf{4.6219}} &
  \multicolumn{1}{c|}{1.7275} &
  \multicolumn{1}{c|}{\textbf{0.69268}} &
  \multicolumn{1}{c|}{1.0267} &
  \multicolumn{1}{c|}{\textbf{1.0142}} \\
 &
  \multicolumn{1}{l|}{\textbf{}} &
  \textbf{} &
  \multicolumn{1}{l|}{\textbf{}} &
  \textbf{} &
  \multicolumn{1}{l|}{\textbf{}} &
  \textbf{} &
  \multicolumn{1}{l|}{\textbf{}} &
  \textbf{} \\ \hline
 &
  \multicolumn{1}{l|}{\textbf{}} &
  \textbf{} &
  \multicolumn{1}{l|}{\textbf{}} &
  \textbf{} &
  \multicolumn{1}{l|}{\textbf{}} &
  \textbf{} &
  \multicolumn{1}{l|}{\textbf{}} &
  \textbf{} \\
\multicolumn{1}{|c|}{$R^2$-Score} &
  \multicolumn{1}{c|}{0.84743} &
  \multicolumn{1}{c|}{\textbf{0.9022}} &
  \multicolumn{1}{c|}{0.87349} &
  \multicolumn{1}{c|}{\textbf{0.93372}} &
  \multicolumn{1}{c|}{0.65830} &
  \multicolumn{1}{c|}{\textbf{0.9450}} &
  \multicolumn{1}{c|}{0.74821} &
  \multicolumn{1}{c|}{\textbf{0.75427}} \\
 &
  \multicolumn{1}{l|}{} &
   &
  \multicolumn{1}{l|}{} &
   &
  \multicolumn{1}{l|}{} &
   &
  \multicolumn{1}{l|}{} &
   \\ \hline
\end{tabular}%
}
\end{table*}

\begin{table*}[]
\centering
\caption{MAE, RMSE and $R^2$-Score of LSTM-based mobile traffic prediction and LLM-based prediction while using training data from different scenarios}
\label{table8}
\resizebox{1.8\columnwidth}{!}{%
\begin{tabular}{cccccccccc}
\multicolumn{10}{c}{The test data are from the "Downloading files - Driving" scenario.} \\
 &
   &
   &
   &
   &
   &
   &
   &
   &
   \\ \hline
\multicolumn{1}{|c|}{} &
   &
  \multicolumn{1}{c|}{} &
   &
  \multicolumn{1}{c|}{} &
   &
  \multicolumn{1}{c|}{} &
   &
  \multicolumn{1}{c|}{} &
  \multicolumn{1}{c|}{} \\
\multicolumn{1}{|c|}{\begin{tabular}[c]{@{}c@{}}Training data\\ is from:\end{tabular}} &
  \multicolumn{2}{c|}{\begin{tabular}[c]{@{}c@{}}Downloading\\ files - Driving\end{tabular}} &
  \multicolumn{2}{c|}{\begin{tabular}[c]{@{}c@{}}Downloading\\ files - Static\end{tabular}} &
  \multicolumn{2}{c|}{\begin{tabular}[c]{@{}c@{}}Watching \\ Amazon videos\\  - Driving\end{tabular}} &
  \multicolumn{2}{c|}{\begin{tabular}[c]{@{}c@{}}Watching \\ Netflix videos\\  - Driving\end{tabular}} &
  \multicolumn{1}{c|}{} \\
\multicolumn{1}{|c|}{} &
   &
  \multicolumn{1}{c|}{} &
   &
  \multicolumn{1}{c|}{} &
   &
  \multicolumn{1}{c|}{} &
   &
  \multicolumn{1}{c|}{} &
  \multicolumn{1}{c|}{\multirow{2}{*}{\begin{tabular}[c]{@{}c@{}}Zero-shot \\ prediction\end{tabular}}} \\ \cline{1-9}
\multicolumn{1}{|c|}{} &
  \multicolumn{1}{c|}{} &
  \multicolumn{1}{c|}{} &
  \multicolumn{1}{c|}{} &
  \multicolumn{1}{c|}{} &
  \multicolumn{1}{c|}{} &
  \multicolumn{1}{c|}{} &
  \multicolumn{1}{c|}{} &
  \multicolumn{1}{c|}{} &
  \multicolumn{1}{c|}{} \\
\multicolumn{1}{|c|}{Methods} &
  \multicolumn{1}{c|}{\begin{tabular}[c]{@{}c@{}}LLM-based\\ prediction\end{tabular}} &
  \multicolumn{1}{c|}{\begin{tabular}[c]{@{}c@{}}LSTM-based\\ prediction\end{tabular}} &
  \multicolumn{1}{c|}{\begin{tabular}[c]{@{}c@{}}LLM-based\\ prediction\end{tabular}} &
  \multicolumn{1}{c|}{\begin{tabular}[c]{@{}c@{}}LSTM-based\\ prediction\end{tabular}} &
  \multicolumn{1}{c|}{\begin{tabular}[c]{@{}c@{}}LLM-based\\ prediction\end{tabular}} &
  \multicolumn{1}{c|}{\begin{tabular}[c]{@{}c@{}}LSTM-based\\ prediction\end{tabular}} &
  \multicolumn{1}{c|}{\begin{tabular}[c]{@{}c@{}}LLM-based\\ prediction\end{tabular}} &
  \multicolumn{1}{c|}{\begin{tabular}[c]{@{}c@{}}LSTM-based\\ prediction\end{tabular}} &
  \multicolumn{1}{c|}{} \\
\multicolumn{1}{|c|}{} &
  \multicolumn{1}{c|}{} &
  \multicolumn{1}{c|}{} &
  \multicolumn{1}{c|}{} &
  \multicolumn{1}{c|}{} &
  \multicolumn{1}{c|}{} &
  \multicolumn{1}{c|}{} &
  \multicolumn{1}{c|}{} &
  \multicolumn{1}{c|}{} &
  \multicolumn{1}{c|}{} \\ \hline
\multicolumn{1}{|c|}{} &
  \multicolumn{1}{c|}{\textbf{}} &
  \multicolumn{1}{c|}{} &
  \multicolumn{1}{c|}{} &
  \multicolumn{1}{c|}{} &
  \multicolumn{1}{c|}{} &
  \multicolumn{1}{c|}{} &
  \multicolumn{1}{c|}{} &
  \multicolumn{1}{c|}{} &
  \multicolumn{1}{c|}{} \\
\multicolumn{1}{|c|}{MAE} &
  \multicolumn{1}{c|}{9.4913} &
  \multicolumn{1}{c|}{\textbf{7.3187}} &
  \multicolumn{1}{c|}{9.8671} &
  \multicolumn{1}{c|}{\textbf{6.9010}} &
  \multicolumn{1}{c|}{\textbf{11.027}} &
  \multicolumn{1}{c|}{20.380} &
  \multicolumn{1}{c|}{\textbf{11.246}} &
  \multicolumn{1}{c|}{23.974} &
  \multicolumn{1}{c|}{13.816} \\
\multicolumn{1}{|c|}{} &
  \multicolumn{1}{c|}{} &
  \multicolumn{1}{c|}{\textbf{}} &
  \multicolumn{1}{c|}{} &
  \multicolumn{1}{c|}{\textbf{}} &
  \multicolumn{1}{c|}{} &
  \multicolumn{1}{c|}{} &
  \multicolumn{1}{c|}{} &
  \multicolumn{1}{c|}{} &
  \multicolumn{1}{c|}{} \\ \hline
\multicolumn{1}{|c|}{} &
  \multicolumn{1}{c|}{} &
  \multicolumn{1}{c|}{\textbf{}} &
  \multicolumn{1}{c|}{} &
  \multicolumn{1}{c|}{\textbf{}} &
  \multicolumn{1}{c|}{} &
  \multicolumn{1}{c|}{} &
  \multicolumn{1}{c|}{} &
  \multicolumn{1}{c|}{} &
  \multicolumn{1}{c|}{} \\
\multicolumn{1}{|c|}{RMSE} &
  \multicolumn{1}{c|}{21.827} &
  \multicolumn{1}{c|}{\textbf{17.469}} &
  \multicolumn{1}{c|}{22.163} &
  \multicolumn{1}{c|}{\textbf{16.104}} &
  \multicolumn{1}{c|}{\textbf{23.457}} &
  \multicolumn{1}{c|}{51.640} &
  \multicolumn{1}{c|}{\textbf{23.149}} &
  \multicolumn{1}{c|}{56.984} &
  \multicolumn{1}{c|}{39.585} \\
\multicolumn{1}{|c|}{} &
  \multicolumn{1}{c|}{} &
  \multicolumn{1}{c|}{\textbf{}} &
  \multicolumn{1}{c|}{} &
  \multicolumn{1}{c|}{\textbf{}} &
  \multicolumn{1}{c|}{} &
  \multicolumn{1}{c|}{} &
  \multicolumn{1}{c|}{} &
  \multicolumn{1}{c|}{} &
  \multicolumn{1}{c|}{} \\ \hline
\multicolumn{1}{|c|}{} &
  \multicolumn{1}{c|}{} &
  \multicolumn{1}{c|}{\textbf{}} &
  \multicolumn{1}{c|}{} &
  \multicolumn{1}{c|}{\textbf{}} &
  \multicolumn{1}{c|}{} &
  \multicolumn{1}{c|}{} &
  \multicolumn{1}{c|}{} &
  \multicolumn{1}{c|}{} &
  \multicolumn{1}{c|}{} \\
\multicolumn{1}{|c|}{$R^2$-Score} &
  \multicolumn{1}{c|}{0.84744} &
  \multicolumn{1}{c|}{\textbf{0.90228}} &
  \multicolumn{1}{c|}{0.84271} &
  \multicolumn{1}{c|}{\textbf{0.91695}} &
  \multicolumn{1}{c|}{\textbf{0.82381}} &
  \multicolumn{1}{c|}{0.14606} &
  \multicolumn{1}{c|}{\textbf{0.82840}} &
  \multicolumn{1}{c|}{0.0398} &
  \multicolumn{1}{c|}{0.49822} \\
\multicolumn{1}{|c|}{} &
  \multicolumn{1}{c|}{\textbf{}} &
  \multicolumn{1}{c|}{} &
  \multicolumn{1}{c|}{} &
  \multicolumn{1}{c|}{} &
  \multicolumn{1}{c|}{} &
  \multicolumn{1}{c|}{} &
  \multicolumn{1}{c|}{} &
  \multicolumn{1}{c|}{} &
  \multicolumn{1}{c|}{} \\ \hline
\end{tabular}%
}
\end{table*}

In this subsection, we compare the performance of ICL-enabled LLM-based mobile traffic prediction and LSTM-based mobile traffic prediction. Table \ref{table6} shows the MAE, RMSE, and $R^2$-Scores of LLM-based prediction and LSTM-based prediction in different application scenarios. As can be observed, in most cases, the LSTM-based method achieves better performance than the LLM-based method. This is because the LSTM models are trained based on a huge training dataset, which makes them better capture historical traffic patterns. However, the superiority of the LSTM model comes at the cost of a large number of training computations and a rich training dataset. This will limit the use of LSTM models in some practical application scenarios. Meanwhile, LSTM models trained in one specific scenario cannot be applied to the prediction tasks in other scenarios, whereas LLM-based methods can work in a one-fits-all manner. In the following, we provide some other numerical results to further illustrate this point.

We perform an experiment using the training data and test data from different application scenarios. For the LLM-based method, the test data are from the "Downloading files - Driving" scenario and the ICL demonstrations are selected from other application scenarios. The numerical results are shown in Table \ref{table8}. According to the numerical results, if the ICL demonstrations are also selected from the “downloading files” scenario, the MAE and RMSE of the traffic prediction are relatively low and the $R^2$-Scores are relatively high. If the ICL demonstrations are selected from other scenarios such as watching Amazon videos or watching Netflix videos, the prediction performance is worse but it is still much better than the zero prediction results. This validates that the ICL demonstrations from other scenarios can also provide useful information to the LLMs and are beneficial to the task. For example, the ICL demonstrations from different scenarios may indicate the relationship between the traffic bitrate and the contextual information. With this information and the base knowledge, the pre-trained LLMs are still able to produce relatively satisfactory predictive results and achieve $R^2$-Scores above 0.8. These numerical results show that the ICL demonstrations and test instances are not necessarily from the same application scenario. As a result, the proposed approach naturally adapts to dynamic network environments in a plug-and-play manner.

Similarly, for the LSTM-based method, the test data are from the "Downloading files - Driving" scenario and the training data are from other application scenarios. The numerical results are shown in Table \ref{table8}. According to the numerical results, if both the test data and the training data come from scenarios with the same activities, which is downloading files in our case, then the LSTM model still maintains good performance. However, if the test data and the training data come from scenarios with different activities, then the LSTM model will have a much worse performance compared with the LLM-based method. These simulation results indicate that LSTMs perform well in capturing scenario-specific data patterns, and they can achieve high accuracy when trained on abundant data. However, their generalization ability across scenarios is limited, especially when scenario-specific data is scarce. These numerical results further corroborate the limitations of the LSTM-based method.

Beyond LSTM-based approaches, transformer-based approaches are also frequently used for traffic prediction. These approaches build upon the foundations of LSTM-based approaches and incorporate more complex architectures and attention mechanisms to further enhance the predictive accuracy. According to the simulation results of LSTM-based approaches they may also achieve higher prediction accuracy than our method. However, transformer models have similar limitations as LSTM-based methods. They typically require retraining for each new deployment scenario and depend heavily on the availability of scenario-specific labeled data. In contrast, the proposed method offers better generalization across different scenarios without the need for scenario-specific data and retraining, which is a critical advantage in real-world applications.

A potential direction is to integrate LLMs with LSTM models or transformer-based models during online inference. Given the strong generalization capabilities of LLMs, they can be utilized in the initial iterations when the prediction task is changed to a new scenario with limited scenario-specific data. As more scenario-specific data becomes available, the LSTM models or transformer-based models can be trained and employed to enhance the prediction accuracy.

\begin{figure}[t]
    \centering
    \subfigure[Traffic prediction results in the case of driving, and downloading files.]{
        \includegraphics[width=5cm,height=3.3cm]{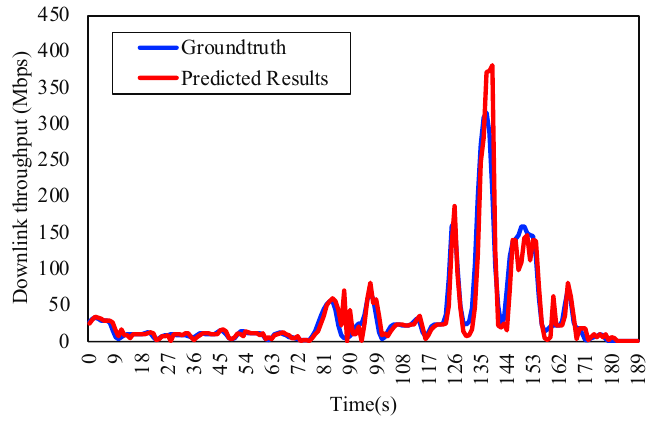}
    }
    \subfigure[Traffic prediction results in the case of static, and downloading files.]{
	\includegraphics[width= 5cm,height=3.3cm]{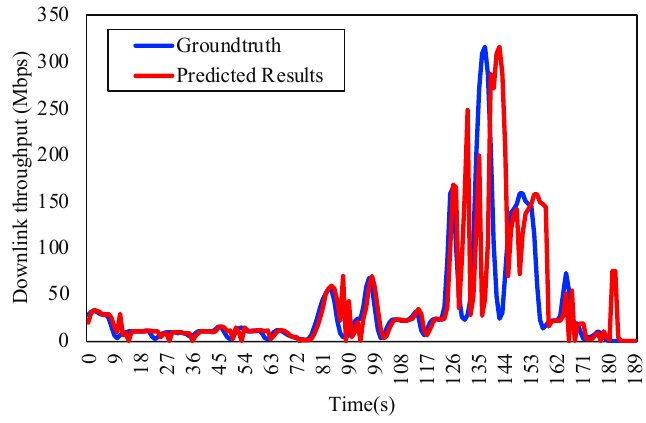}
    }
    \caption{Traffic prediction with selected ICL demonstrations using PHI-3-mini.}
    \label{fig5}

\end{figure}

\begin{table}[]
\centering
\caption{MAE, RMSE and $R^2$-Score of mobile traffic prediction using PHI-3-mini.}
\label{table5}
\resizebox{0.8\columnwidth}{!}{%
\begin{tabular}{|l|l|l|l|l|}
\hline
                               &                               &                              &                                       &                              \\
\multicolumn{1}{|c|}{} &
  \multicolumn{1}{c|}{\begin{tabular}[c]{@{}c@{}}Zero-shot\\ prediction\end{tabular}} &
  \multicolumn{1}{c|}{\begin{tabular}[c]{@{}c@{}}Proposed \\ one-demo \\ prediction\end{tabular}} &
  \multicolumn{1}{c|}{\begin{tabular}[c]{@{}c@{}}Proposed \\ two-demo \\ prediction\end{tabular}} &
  \multicolumn{1}{c|}{\begin{tabular}[c]{@{}c@{}}Full\\ ICL\end{tabular}} \\
                               &                               &                              &                                       &                              \\ \hline
                               &                               &                              &                                       & \textbf{}                    \\
\multicolumn{1}{|c|}{MAE}      & \multicolumn{1}{c|}{23.990}   & \multicolumn{1}{c|}{11.824}  & \multicolumn{1}{c|}{\textbf{9.4348}}  & \multicolumn{1}{c|}{10.112}  \\
                               & \textbf{}                     &                              &                                       & \textbf{}                    \\ \hline
                               & \textbf{}                     &                              &                                       & \textbf{}                    \\
\multicolumn{1}{|c|}{RMSE}      & \multicolumn{1}{c|}{58.730}   & \multicolumn{1}{c|}{25.833}  & \multicolumn{1}{c|}{\textbf{21.088}}  & \multicolumn{1}{c|}{22.342}  \\
                               & \textbf{}                     &                              &                                       & \textbf{}                    \\ \hline
                               & \textbf{}                     &                              &                                       & \textbf{}                    \\
\multicolumn{1}{|c|}{$R^2$-Score} & \multicolumn{1}{c|}{-0.10452} & \multicolumn{1}{c|}{0.78630} & \multicolumn{1}{c|}{\textbf{0.85759}} & \multicolumn{1}{c|}{0.84016} \\
                               &                               &                              & \textbf{}                             & \textbf{}                    \\ \hline
\end{tabular}%
}
\end{table}

\subsection{Numerical Results with PHI-3-mini model}
We also provide some traffic prediction results with the PHI-3-mini model in Fig. \ref{fig5} and Table \ref{table5}. As can be observed from the results, the proposed ICL demonstration selection method is also applicable to the PHI-3-mini model. Specifically, the zero-shot prediction results of the PHI-3-mini model are much worse than that of the PHI-3-medium model. In the zero-shot prediction case, the PHI-3-mini model yields a negative $R^2$-Score, which indicates that the prediction results are worse than simply using the mean of the target values as a predictor. However, by using ICL with selected demonstrations, PHI-3-mini requires only about one-fourth of the model parameters to achieve predictive performance comparable to that of the PHI-3-medium model.

This study mainly focuses on PHI-3-mini and PHI-3-medium. However, the proposed method is inherently model-agnostic since the architecture of the LLM is treated as a black box. As a result, it is also applicable to other LLMs with different architectures. According to the comparison between PHI-3-mini and PHI-3-medium, we can expect the similar trend if larger LLMs, such as GPT-4 and Llama-4, are used. It is also mentioned in some academic papers that increasing model size does not always improve ICL performance and may even worsen it under certain conditions [35]. So the larger models like GPT-4 and Llama-4 are expected to lead to better performance in traffic prediction tasks without ICL. However, their performance under our proposed ICL framework is not necessarily better than that of smaller models.

On the other hand, using larger LLMs in wireless networks bring several limitations. Large models typically require significantly more memory and computation resources. For mobile traffic prediction, which may need to run in latency-sensitive and resource-constrained environments, using such larger models is kind of impractical.

\section{Conclusions}
Mobile traffic prediction provides support for wireless communication network optimization and decision-making. However, real-world wireless network traffic data are highly nonlinear and complex, which makes mobile traffic prediction a challenging task. In this work, we designed an LLM-enabled mobile traffic prediction framework and proposed a two-step ICL demonstration selection scheme. This framework can accurately and stably predict the traffic in various scenarios. Specifically, we proposed two rules—effectiveness and informativeness—to evaluate the ICL demonstrations, based on a thorough theoretical analysis. Effectiveness means that the selected ICL demonstrations should be helpful for the test case. Informativeness means that the selected ICL examples should provide information that the original pre-trained LLM does not have. The superiority of the proposed framework is demonstrated on a real-world 5G dataset. We found that ICL performs better than zero-shot prediction and with the proposed demonstration selection method, two demonstrations are sufficient for accurate prediction. ICL also helps bridge the gap between LLMs with different capabilities, such as PHI-3 mini and PHI-3 medium model. In addition, LSTM models perform better when testing on the same application scenario and they are usually much smaller than LLMs and have faster inference time. However, the proposed method is much better than LSTM in generalizing over new scenarios.

For future work, hybrid model architectures can be explored to further leverage the advantages of both models. For instance, LLMs can generate generalized coarse-grained predictions to constrain the range of LSTM outputs, or, the coarse-grained predictions can be used as additional features that enhance LSTM performance. Meanwhile, LSTMs can focus on fine-grained predictions for the given scenarios. In addition, in this work, we only focus on ICL. Future work could explore hybrid methods combining our selection strategy with other advanced prompt engineering techniques, such as retrieval-augmented generation and chain-of-thought, to further enhance the prediction performance.

Moreover, there are still scenarios where ICL selection may fail to improve performance. For example, if a test input falls into a sparse region of the feature space, and the available demonstrations are limited, even the most similar ones may still differ significantly in distribution or semantics. To mitigate this limitation, one potential direction is to incorporate a confidence-aware selection mechanism to filter out ICL demonstrations whose contribution is uncertain or potentially harmful. We plan to investigate these ideas in future work.

\section*{Acknowledgment}
This work has been supported by MITACS and Ericsson
Canada, and NSERC Canada Research Chairs program.

\normalem
\bibliographystyle{IEEEtran}
\bibliography{reference}

\vspace{-40pt}

\end{document}